\documentclass[a4paper,fleqn]{cas-dc}
\usepackage[numbers]{natbib}
\usepackage[utf8]{inputenc}
\usepackage{tabularray}
\usepackage{cleveref}
\usepackage{caption}
\usepackage{hyperref} 
\usepackage{algorithmic}
\usepackage{algorithm}
\usepackage{float}  
\usepackage{subfig}
\usepackage{graphicx}

\def\tsc#1{\csdef{#1}{\textsc{\lowercase{#1}}\xspace}}
\tsc{WGM}
\tsc{QE}
\tsc{EP}
\tsc{PMS}
\tsc{BEC}
\tsc{DE}

\begin{document}
\let\WriteBookmarks\relax
\def\floatpagepagefraction{1}
\def\textpagefraction{.001}
\let\printorcid\relax 
\captionsetup[figure]{labelfont={bf}, labelformat={default}, labelsep=period, name={Fig.}}

\shorttitle{FE-MCFormer: A novel time–frequency interpretable fault diagnosis architecture}

\shortauthors{Yuhan Yuan et~al.}

\title [mode = title]{FE-MCFormer: a novel time–frequency interpretable architecture for machinery fault diagnosis under strong noise environments}                      

\author[1]{Yuhan Yuan}[style=chinese]
\ead{yuhanyuan@mail.dlut.edu.cn}
\credit{Conceptualization, Data curation, Formal analysis, Investigation, Methodology, Software, Validation, Writing – original draft}

\affiliation[1]{organization={School of Energy and Power Engineering},
    addressline={Dalian University of Technology}, 
    city={Dalian},
    postcode={116024}, 
    country={China}}

\author[1,2,3]{Xiaomo Jiang}[style=chinese]
\cormark[1]
\ead{xiaomojiang2019@dlut.edu.cn}
\credit{Funding acquisition, Project administration, Resources, Supervision, Writing – review & editing}
\affiliation[2]{organization={State Key Lab of Structural Analysis, Optimization and CAE Software for Industrial Equipment},
    addressline={Dalian University of Technology}, 
    city={Dalian},
    postcode={116024}, 
    country={China}}
    
\affiliation[3]{organization={Provincial Key Lab of Digital Twin for Industrial Equipment},
    addressline={Dalian University of Technology}, 
    city={Dalian},
    postcode={116024}, 
    country={China}}
    
\author[1]{Haibin Yang}[style=chinese]
\ead{yanghaibin@mail.dlut.edu.cn}
\credit{Software, Data curation, Validation}

\author[1]{Haixin Zhao}[style=chinese]
\ead{zhaohaixin@dlut.edu.cn}
\credit{Data curation, Validation, Writing – review & editing}

\author[1]{Shengbo Wang}[style=chinese]
\ead{s2013566862@mail.dlut.edu.cn}
\credit{Software, Data curation, Validation}

\author[4]{Xueyu Cheng}[style=chinese]
\ead{xueyucheng@clayton.edu}
\credit{Validation, Supervision, Writing – review & editing}
\affiliation[4]{organization={College of Arts and Sciences, Clayton State University},
    city={Morrow},
    state = {GA},
    postcode={30260},
    country = {USA}}
    
\author[1]{Jigang Meng}[style=chinese]
\ead{mengjigang@mail.dlut.edu.cn}
\credit{Software, Investigation, Data curation}

\cortext[cor1]{Corresponding author at: State Key Lab of Structural Analysis, Optimization and CAE Software for Industrial Equipment, Dalian University of Technology, Dalian, 116024, China.}

\begin{abstract}
Interpretable fault diagnosis (FD) plays a critical role in industrial manufacturing, as it improves human–machine understanding and operational efficiency. However, harsh operating environments often introduce strong background interference or noise, which weakens the discriminative capability and interpretability of existing FD methods. To address this issue, this paper proposes FE-MCFormer, a time–frequency fusion framework for robust and time–frequency interpretable fault diagnosis under strong noise conditions. A frequency adaptive learning layer (FALL) is developed to perform learnable spectral reconstruction, which explicitly suppresses noise-dominated frequency responses while preserving fault-sensitive harmonic structures. Furthermore, a multiscale time–frequency fusion (MSTFF) architecture is designed to jointly capture localized impulsive characteristics and structured global spectral interactions. Extensive experiments on a rolling bearing dataset and a real-world centrifugal compressor dataset demonstrate that the proposed method achieves stable and interpretable diagnostic performance under severe noise environments down to \(-10\) dB SNR. The results indicate that FE-MCFormer provides an effective framework for turbomachinery fault diagnosis in complex noisy environments.
\end{abstract}

\begin{highlights}
\item A novel network architecture is proposed for time–frequency interpretable machinery fault diagnosis under strong noise conditions.
\item A frequency adaptive learning layer is proposed for first layer interpretable diagnosis.
\item A multiscale time–frequency fusion module is designed to enhance both the noise robustness and interpretability of the proposed method.
\item The effectiveness and interpretability of the proposed model are validated on a bearing dataset and a centrifugal compressor dataset.
\end{highlights}


\maketitle

\section{Introduction}
\label{sec:introduction}
\subsection{Background and Motivation}
Rotating machinery plays a critical role in modern industrial systems in energy, manufacturing, and process industries \cite{lei2020applications, ahmad2025consistency}. The operational reliability of these machines directly affects production safety, energy efficiency, and maintenance costs. Unexpected failures of rotating components may lead to catastrophic system breakdowns and substantial economic losses \cite{ZHAO2026104189}. Therefore, accurate and robust fault diagnosis (FD) has become an essential requirement for intelligent condition monitoring and predictive maintenance of rotating machinery.

Fault diagnosis methods based on traditional signal processing extract handcrafted features from time, frequency, and time–frequency domains. For example, Jiang et al. \cite{jiang_time-frequency_2023} developed a time–frequency fusion approach of the spectral amplitude to capture the characteristic frequencies of vibration signals. Chen et al. \cite{chen_feature_2023} developed a hierarchical improved envelope spectrum entropy method to extract intrinsic fault features of mechanical vibration signals. Kou et al. \cite{kou_application_2020} applied a novel empirical mode decomposition (EMD) approach with empirical entropy (EE) to extract features and validated the method on mine hoist sheave bearing fault detection. Zhou et al. \cite{zhou_novel_2023} proposed a continuous hierarchical fractional range entropy approach for machine FD in noisy environments. Although these approaches have achieved notable success in laboratory settings, they usually require extensive expert knowledge and show limited robustness under strong noise environments or complex industrial operating conditions. This restricts their scalability and applicability in practical industrial informatics systems.

Compared with traditional signal processing approaches, machine learning (ML) methods can extract more valuable fault characteristics from large historical datasets, making them a promising approach for machinery FD. For example, Ren et al. \cite{ren_meta-learning_2024} introduced an unsupervised across-tasks meta-learning strategy to enable few-shot FD under varying working conditions. Zhang et al. \cite{zhang_semi-supervised_2023} developed a representation-based transfer learning approach for enhancing few-shot bearing fault classification. However, machine learning approaches exhibit limitations in modeling high-dimensional and nonlinear variations, which can hinder their effectiveness in complex industrial scenarios.

With the rapid development of artificial intelligence and data-driven techniques, data-driven FD methods have gained widespread attention. By automatically extracting discriminative representations from raw signals, data-driven models reduce the reliance on manual feature engineering and demonstrate superior performance under certain conditions. For example, Shao et al. \cite{he2026hybrid} developed a novel multiscale network for few-shot rotating machinery fault diagnosis. Jia et al. \cite{jia_multiscale_2022} developed a multiscale attention convolutional network to reduce noise from vibrational signals. Chen et al. \cite{CHEN2022501} proposed a multi-channel calibrated transformer for few-shot bearing fault diagnosis. Liao et al. \cite{LIAO2025111750} proposed a physics-informed blind deconvolution network for robust bearing fault diagnosis. Chen et al. \cite{CHEN2026130747} developed a multi-scale convolutional network for bearing fault diagnosis under noisy conditions. Wang et al. \cite{WANG2025112889} introduced a novel network, which integrates 1D and 2D feature information for efficient feature extraction and classification. Although the aforementioned methods have significantly advanced machinery health monitoring, the internal mechanisms underlying their diagnostic decisions are rarely investigated. This lack of interpretability may weaken the reliability and practical applicability of intelligent operation and maintenance systems in real industrial scenarios.

To address this limitation, recent studies have introduced interpretability capability into FD models. These methods are typically categorized into time-domain and frequency-domain interpretability. Specifically, time-domain interpretability is usually achieved by guiding the model to focus on critical temporal segments, such as fault-induced impulsive regions in signals. In comparison, frequency-domain methods generally integrate prior knowledge from signal processing to associate learned representations with meaningful spectral components. 

Time-domain interpretable methods usually incorporate prior constraints or attention mechanisms to guide networks toward informative temporal regions, such as transient impulses induced by mechanical faults. For example, Li et al. \cite{li_variational_2024} proposed an interpretable variational transformer with a Dirichlet prior for fault diagnosis and analyzed its interpretability through attention weight visualization. Han et al. \cite{YUAN2025322} developed a local attention network for degradation feature extraction of industrial robots, further explaining how the attention mechanism highlights degradation-related features. Zhang et al. \cite{zhang2025wd} proposed a KAN-Transformer-based rotating machinery FD framework and conducted weight heatmap analysis to reveal the internal operating mechanism of the network. Although these methods \cite{li_variational_2024, YUAN2025322, zhang2025wd} demonstrate promising diagnostic performance by extracting interpretable time-domain features, their applicability may be limited when fault modes do not exhibit pronounced time-domain impulsive signatures. For instance, in sliding bearings of large-scale centrifugal compressor units, fault-related information is often more clearly reflected in spectral variations and harmonic components. 

Methods focusing on frequency-domain interpretability usually embed prior spectral knowledge or physical constraints into neural networks. The interpretability of these methods is commonly examined by visualizing the learned spectral weights or reconstructed frequency components. For example, Liu et al. \cite{LIU2025111271} proposed a Transformer-based interpretable network for few-shot bearing fault diagnosis. Wang et al.\cite{10870139} proposed a global explainable convolutional neural network for bearing fault diagnosis. Chen et al.\cite{CHEN2024102705} proposed a novel interpretable bearing fault diagnosis method based on convolutional neural network and variational inference. Zhou et al.\cite{10584499} proposed a CNN-LSTM-based interpretable network for rotating machinery fault diagnosis. Although the above methods \cite{LIU2025111271, 10870139, CHEN2024102705, 10584499} show great potential in interpretable fault diagnosis, the aforementioned studies rarely consider the influence of strong noise environments. In practical industrial environments, measured signals are inevitably contaminated by strong noise due to harsh operating conditions, variable loads, electromagnetic interference, and complex transmission paths. Under such circumstances, fault-related impulsive features are often weak and easily submerged by noise, making reliable fault diagnosis particularly challenging. In summary, current interpretable FD methods still face the following challenges:

\begin{enumerate}[1)]
\item Although current interpretability studies can extract fault-related features in a single time or frequency domain, time–frequency interpretability remains rarely investigated. As a result, this limitation restricts their generality and applicability in practical industrial fault diagnosis.

\item The aforementioned studies have achieved promising performance under ideal experimental conditions. However, real industrial environments are often contaminated by strong noise. This poses a significant challenge to the reliability of interpretable fault diagnosis models in real-world production scenarios.

\end{enumerate}

In order to address the existing challenges, this work introduces a time–frequency interpretable and noise-robust fault diagnosis framework named FE-MCFormer (Frequency-Enhanced Multiscale convformer). The proposed method can extract interpretable components from both time and frequency domains, thereby achieving time–frequency interpretability from both the first-layer and deep feature representations under strong noise conditions. Specifically, the proposed method can adaptively learn hierarchical frequency representations from input signals through a novel network architecture. This architecture can adaptively guide the model to focus on fault-sensitive frequency-domain components, enhancing its robustness against noise interference. Moreover, local attention mechanisms are employed to extract interpretable temporal components, further improving the model’s interpretability and diagnostic accuracy. 

The main contributions of this paper are summarized as follows:
\begin{enumerate}
\item A novel network architecture, named FE-MCFormer, is developed for time–frequency interpretable fault diagnosis of turbomachinery under strong noise. The time–frequency features of measurement signals are captured in the proposed framework through the adept integration of the time–frequency interpretable features.

\item A frequency adaptive learning layer (FALL) is introduced to adaptively encode global spectral information and suppress noise-dominated frequency components in the first layer of the model. This layer enhances both the noise robustness and interpretability of the proposed method.

\item A multiscale time–frequency fusion module is designed by combining multiscale attention and time–frequency fusion networks. This design further enhances the noise robustness and interpretability of the model.

\item The effectiveness and interpretability of FE-MCFormer are validated using a benchmark laboratory dataset and a real-world centrifugal compressor dataset. Extensive experiments, including robustness evaluation under varying signal-to-noise ratios, ablation studies, and interpretability analysis, demonstrate the effectiveness and industrial applicability of the proposed framework.

\end{enumerate}

\subsection{Paper Organization}
The remainder of this paper is given as follows. Section \ref{Methodology} describes the network architecture of the proposed FE-MCFormer and details its key components, including the frequency adaptive learning layer and the multiscale time–frequency fusion module. Section \ref{Experimental Verification} presents the experimental setup, datasets, and comparative evaluation results, followed by robustness analysis, ablation studies, and interpretability visualization. Section \ref{Conclusion} concludes the paper.

\section{Methodology} \label{Methodology}

\begin{figure}
    \centering
    \includegraphics[width=1\linewidth]{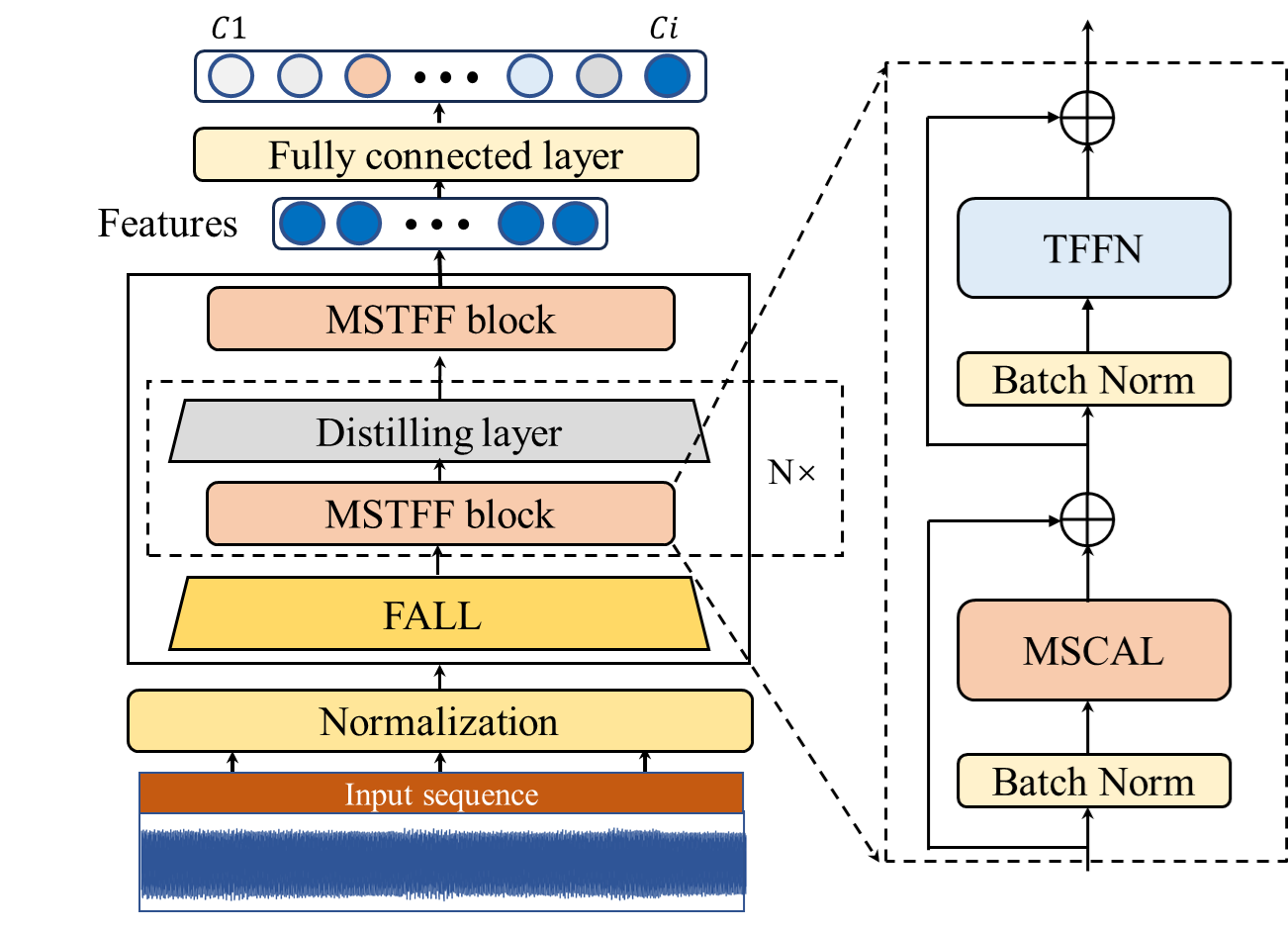}
    \caption{Network architecture of the proposed FE-MCFormer.}
    \label{fig.network}
\end{figure}

This section details the proposed FE-MCFormer framework, designed to provide robust and interpretable fault diagnosis for rotating machinery in harsh industrial environments. The architecture, illustrated in Fig. \ref{fig.network}, integrates a frequency adaptive learning layer (FALL), multiscale time–frequency fusion (MSTFF) modules, and three distillation layers to balance feature extraction performance with computational efficiency. 

\subsection{Frequency adaptive learning layer (FALL)}

Conventional embedding layers often fail to distinguish between fault-related impulsive features and broadband industrial noise. Therefore, a frequency adaptive learning layer (FALL) is proposed to address this. This layer is applied at the input stage to perform first-layer frequency-domain reconstruction and suppress noise-dominated components.

\subsubsection{Operator-Theoretic Formulation}

Given an input feature map
\[
X \in \mathbb{R}^{B \times C \times L},
\]
a linear projection is first applied:

\begin{equation}
\tilde{X} = \mathrm{Conv}_{1\times k}(X),
\end{equation}

\noindent where \(Conv(\cdot)\) stands for the convolutional operation. 

In this layer, a frequency adaptive learning (FAL) strategy is employed to enhance model robustness and interpretability. Let $\mathcal{F} : \mathcal{H} \rightarrow \mathbb{C}^L$ denote the discrete Fourier transform (DFT), which is unitary:

\begin{equation}
\mathcal{F}^{-1} = \mathcal{F}^{*}.
\end{equation}

By Parseval's theorem,

\begin{equation}
\|x\|_2^2 = \|\mathcal{F}x\|_2^2.
\end{equation}

Let $W \in \mathbb{C}^L$ denote a learnable spectral weight vector. Define the diagonal spectral multiplication operator as follows:

\begin{equation}
\mathcal{M}_W(\hat{x}) = W \odot \hat{x},
\end{equation}

\noindent where $\odot$ denotes element-wise multiplication.

The frequency-adaptive learning operation is defined as:

\begin{equation}
\mathcal{T}_F = I + \gamma \mathcal{F}^{-1} \mathcal{M}_W \mathcal{F},
\end{equation}

\noindent where $I$ is the identity operator and $\gamma \in (0,1)$ is a scaling factor. The embedding becomes:

\begin{equation}
x_r = \mathcal{T}_F(x).
\end{equation}

This demonstrates that FALL acts as a learnable spectral shaping operator that attenuates noise-dominated frequency components while preserving fault-related harmonics. As such, FALL serves as a global encoder, effectively integrating local patterns with frequency-domain global dependencies in an adaptive and differentiable manner. Unlike conventional convolutional filters that implicitly learn frequency responses, FALL explicitly performs learnable spectral shaping, which allows the model to decouple noise-dominated high-frequency components from fault-related harmonic structures.

\subsection{Multiscale Time–Frequency Fusion (MSTFF) Operator}

\begin{figure*}
    \centering
    \includegraphics[
        width=0.85\textwidth,
        height=0.55\textheight,
        keepaspectratio
    ]{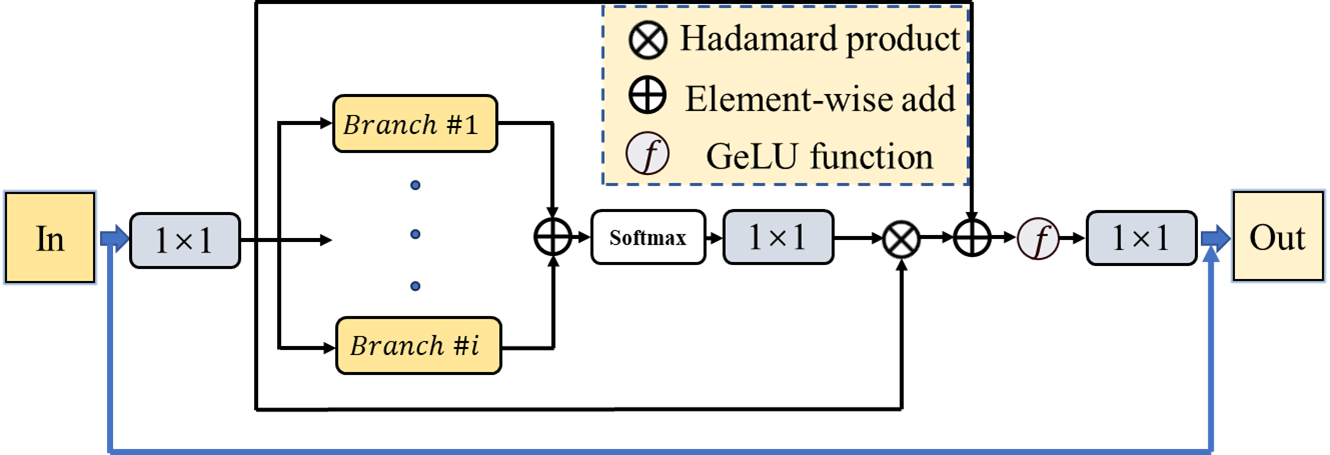}
    \caption{Illustration of the proposed multiscale convolutional attention layer (MSCAL).}
    \label{fig.mscal}
\end{figure*}

\begin{figure*}
    \centering
        \includegraphics[
        width=0.75\textwidth,
        height=0.50\textheight,
        keepaspectratio
    ]{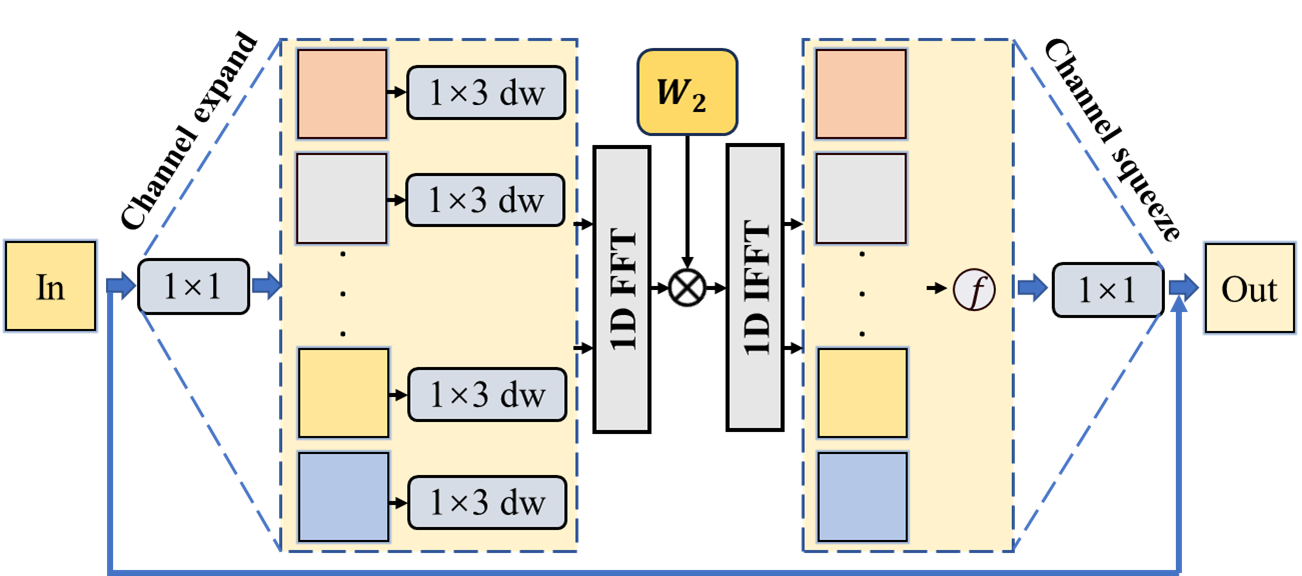}
    \caption{Illustration of the proposed time–frequency fusion network (TFFN).}
    \label{fig.tffn}
\end{figure*}

In this model, a Multiscale Time–Frequency Fusion (MSTFF) operator is designed to capture local impulsive behaviors alongside global spectral structures without the quadratic computational cost of the conventional MHSA mechanism. As illustrated in Fig. \ref{fig.network}, the MSTFF includes a multiscale convolution attention layer (MSCAL) and a time–frequency fusion network (TFFN) with two Batch Normalization layers. 

\subsubsection{MSCAL}

As shown in Fig. \ref{fig.mscal}, MSCAL extracts multiscale temporal features by utilizing parallel branches with $1\times3$ and $1\times5$ convolutional operations. A softmax-based attention mechanism is applied to these branches to adaptively scale features based on their diagnostic relevance. This architecture enables the model to concentrate on localized fault-related features that are often obscured in global sequences. 

Let $X \in \mathbb{R}^{C \times L}$, multiscale temporal operators are constructed as follows:

\begin{equation}
X_k = \mathrm{Conv}_{1\times k}(X), \quad k \in \{3,5\},
\end{equation}

Subsequently, two parallel branches with \(1\times 3\) and \(1\times 5\) convolutional operations are introduced to extract multiscale time-domain features as follows:
\begin{flalign}
Y_2=\sum_{i=1}^{2} Branch_i[X_k],
\label{eq-y2}
\end{flalign}
where \(Branch_i[\cdot]\) denotes the two feature branches with \(1\times 3\) and \(1\times 5\) convolutional operations, which are sensitive to local fault features.

Further, a softmax function is applied to map multiscale features to the value of \([0, 1]\), and another \(1\times 1\) convolution kernel is utilized to fuse the multiscale information. The Hadamard product is performed between fusion features and feature map \(X_k\), and a residual connection is added between \(X_k\) and the output features, given as follows:

\begin{flalign}
Y_3 =\left\{\begin{matrix}Conv_{1\times1}(attn)\otimes X_k + X_k,
 \\attn=softmax(Y_2)
\end{matrix}\right.
\label{eq-y3}
\end{flalign}
where \(\otimes\) denotes the Hadamard product, and \(softmax(\cdot)\) stands for the softmax function. Finally, the GELU function is employed to enhance nonlinear representation capability, and a \(1\times 1\) convolution kernel is introduced as another channel-mixing kernel. Ultimately, the residual structure is employed to prevent gradient vanishing or explosion, given by:
\begin{flalign}
Y_4 = Conv_{1\times1}[GELU(Y_3)] + X,
\end{flalign}

As such, MSCAL acts as a data-adaptive local feature amplifier, enhancing impulsive fault signatures.

\subsubsection{TFFN}

\begin{figure*}
    \centering
    \includegraphics[width=1\linewidth]{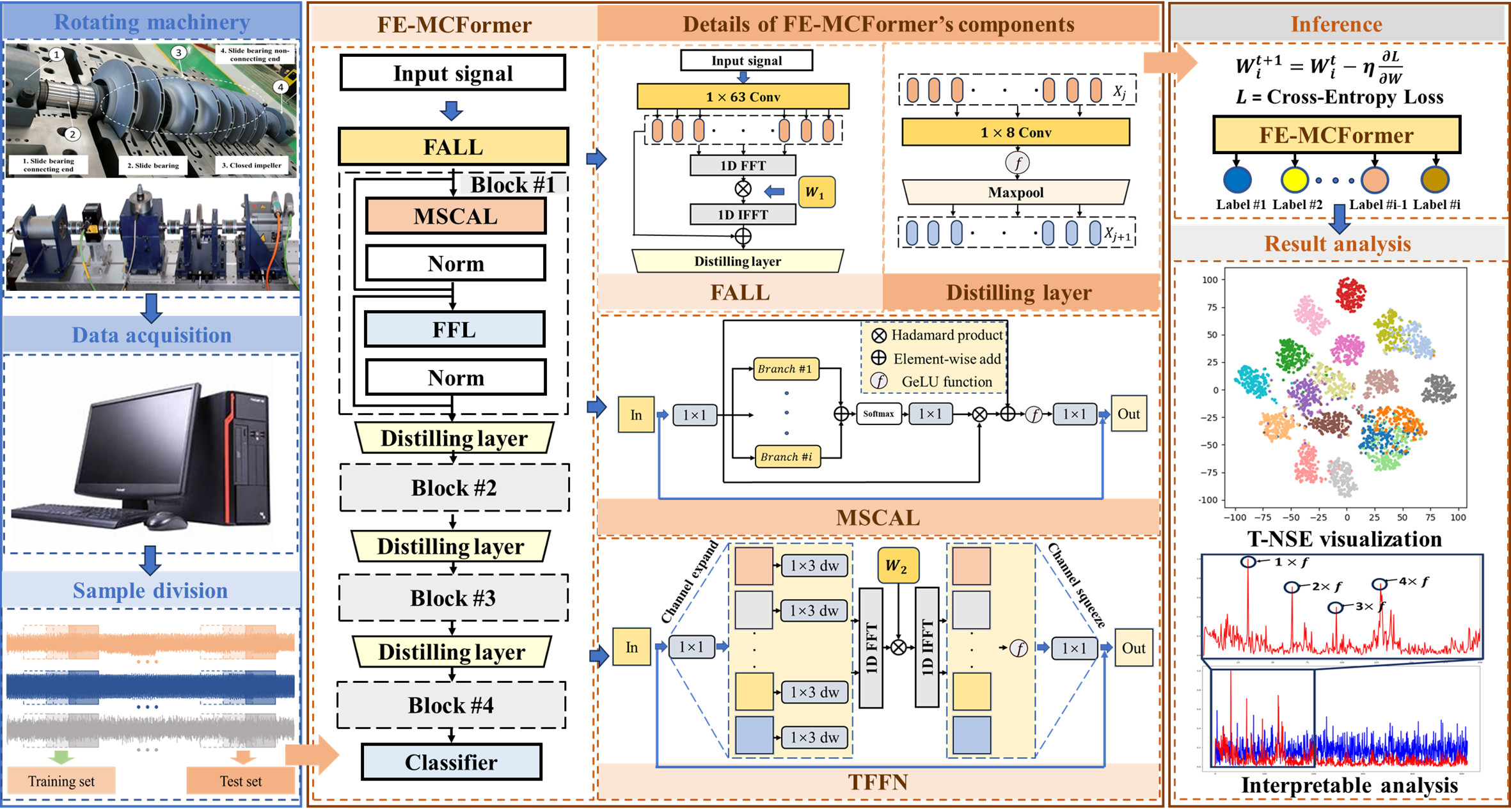}
    \caption{Illustration of the proposed fault diagnosis framework.}
    \label{fig.1}
\end{figure*}

The Time–Frequency Fusion Network (TFFN) is utilized to capture deep features and model time–frequency interactions among high-level semantic representations, as shown in Fig. \ref{fig.tffn}. The TFFN employs a time–frequency fusion adaptive learning operation to filter irrelevant information in the frequency domain and capture local features in the time domain. More specifically, TFFN utilizes a local convolutional operation in the time domain to capture local time-series features and introduces the FAL operation in the frequency domain to model long-range dependencies and enhance nonlinear representation capability.

As shown in Fig. \ref{fig.tffn}, for the input tensor \(X\in \mathbb{R}^{C\times L}\), the TFFN employs a \(1\times 1\) convolutional operation to expand the feature dimensionality, given by:

\begin{flalign}
\bar{X}\in \mathbb{R}^{2C \times L}=Conv_{1 \times 1}^{ex} (X \in \mathbb{R}^{C \times L} ) , 
\end{flalign}
where \(Conv_{1 \times 1}^{ex}(\cdot)\) denotes the \(1 \times 1\) convolution with channel expansion. Subsequently, a \(1 \times 3\) depthwise convolution is applied to capture local patterns, and a GELU activation function is introduced to improve nonlinear representation capability, formulated as:

\begin{flalign}
\dot{X} =GELU[DepthwiseConv_{1 \times 3}^{dw}(\bar{X} ) ],
\end{flalign}
where \(GELU\) denotes the activation function, and \(Conv_{1 \times 3}^{dw}(\cdot)\) represents the \(1\times 3\) depthwise convolutional operation. Furthermore, to obtain frequency-domain representations, the time-domain features are processed using a discrete Fourier transform (DFT), given by

\begin{flalign}
{X}' =F[\dot{X} ],
\end{flalign}
\noindent where \(F[\cdot]\) denotes the DFT operation. Then, a learnable weight \(W\) is introduced to reconstruct the features in the frequency domain, and a scaling factor is utilized to maintain numerical stability, defined by:

\begin{flalign}
\hat{X} =\gamma ({X}'\otimes W),
\label{eq-gamma2}
\end{flalign}
where \(\hat{X}\) represents the reconstructed feature, \(\otimes\) stands for the Hadamard product, and \(\gamma\) denotes the scaling factor. Moreover, an inverse DFT is applied to transform the feature back into the time domain, where a \(1 \times 1\) convolutional operation is applied to transform the output features back into the time-domain, given by:

\begin{flalign}
\check{X} \in R^{C \times L}=Conv_{1 \times1}^{se}[F^{-1}(\hat{X} ) ] ,
\end{flalign}
\noindent where \(Conv_{1 \times1}^{se}(\cdot)\) represents a \(1 \times 1\) convolution with channel reduction, and \(F^{-1}[\cdot]\) stands for the inverse DFT. Finally, the output features are added element-wise to the original feature \(X\) as the final output of TFFN:

\begin{flalign}
X^{r}=X+\check{X},
\end{flalign}
where \(X^r\) stands for the output of TFFN.

\subsubsection{Distilling layer}

To enhance the model's efficiency, a hierarchical distilling layer is positioned between MSTFF blocks. This layer employs a 1-D convolution and a max-pooling operation to halve the temporal dimension, effectively creating a focused global feature map and reducing memory usage for subsequent layers. This distilling procedure from the j-th layer to the (j+1)-th layer is formulated as follows:

\begin{flalign}
&X_{j+1}^{2C\times\frac{L}{2}}
\notag
\\&=MaxPool\left(GELU\left(Conv1d\left(\left[X_{j}^{C\times L}\right]_{MS}\right)\right)\right),
\label{eq:distill}
\end{flalign}

\noindent where \([\cdot]_{MS}\) represents the MSTFF module, and \(Conv1d\) denotes a 1-D convolution kernel with the GELU function. Here, \(C\) and \(L\) represent the number of channels and the sequence length of \(X\). We expand the number of channels by using a \(1\times 5\) convolutional operation and employ a max-pooling layer to downsample \(X\) into half of its temporal length after stacking a layer, which reduces the overall memory usage of the next MSTFF module.

\subsection{Unified Operator Formulation}

The overall FE-MCFormer architecture can be formulated as:

\begin{equation}
f_\theta = \mathcal{C} \circ \left( \prod_{j=1}^{M} \mathcal{D}_j \circ \mathcal{T}_j \right) \circ \mathcal{F},
\label{eq:overall}
\end{equation}

\noindent where $\mathcal{F}$ denotes the FALL spectral encoder, $\mathcal{T}_j$ denotes the $j$-th MSTFF block, $\mathcal{D}_j$ denotes the distillation operator, and $\mathcal{C}$ denotes the classifier.

This formulation unifies spectral filtering, multiscale temporal modeling, global dependency extraction, and hierarchical compression within a computationally efficient framework suitable for industrial fault diagnosis.

\subsection{Computational Complexity Analysis}

In industrial fault diagnosis, vibration signals are typically sampled at high frequency, resulting in long temporal sequences. Therefore, computational efficiency is a critical consideration for real-time monitoring systems.

Let $L$ denote the input length and $C$ the number of channels. We analyze the computational complexity of each major component in FE-MCFormer.

\subsubsection{Complexity of FALL}

FALL consists of a 1-D convolution: $\mathcal{O}(L)$, a forward DFT: $\mathcal{O}(L \log L)$,
 element-wise spectral multiplication: $\mathcal{O}(L)$, and an inverse DFT: $\mathcal{O}(L \log L)$. Thus, the dominant term is

\begin{equation}
\mathcal{O}(FALL)=\mathcal{O}(L \log L)+\mathcal{O}(L)+\mathcal{O}(L \log L)\Rightarrow \mathcal{O}(L \log L)
\end{equation}

\subsubsection{Complexity of MSCAL}

MSCAL mainly consists of two parallel convolutions with kernel sizes $k \in \{3,5\}$ for multiscale feature extraction: $\mathcal{O}(2L)$, and element-wise multiplication: $\mathcal{O}(L)$. Therefore,

\begin{equation}
\mathcal{O}(MSCAL)=\mathcal{O}(2L)+\mathcal{O}(L)\Rightarrow \mathcal{O}(L)
\end{equation}

\subsubsection{Complexity of TFFN}

TFFN mainly consists of a depthwise convolution, spectral transformation, and frequency adaptive learning, including depthwise convolution: $\mathcal{O}(L)$, forward DFT: $\mathcal{O}(L \log L)$, spectral multiplication: $\mathcal{O}(L)$, and inverse DFT: $\mathcal{O}(L \log L)$. Therefore, the overall complexity is determined by

\begin{equation}
\mathcal{O}(TFFN)=\mathcal{O}(L)+\mathcal{O}(L \log L)+\mathcal{O}(L)+\mathcal{O}(L \log L)\Rightarrow \mathcal{O}(L \log L)
\end{equation}

\subsubsection{Complexity of Distillation}

The distillation layer applies convolution and pooling, reducing sequence length from $L_j$ to $L_{j+1} = L_j/2$. Its complexity is

\begin{equation}
\mathcal{O}(L_j).
\end{equation}

Since the sequence length decreases geometrically across layers, the cumulative complexity over $M$ blocks becomes

\begin{equation}
\sum_{j=1}^{M} \mathcal{O}(L_j \log L_j),
\end{equation}
\noindent which remains near-linear in practice. Therefore, the proposed FE-MCFormer achieves near-linear time complexity and reduced computational burden across hierarchical layers, making it well-suited for long vibration sequences and real-time health monitoring in industrial informatics systems.

\section{Results and Discussion} \label{Experimental Verification}

To evaluate the adaptability and practicability of FE-MCFormer and FALL, two case studies are conducted in this section. A benchmark rolling bearing dataset is used in Case 1, while the actual measurements from a real-world centrifugal compressor are utilized in Case 2.

\begin{table*}
\centering
\caption{Network structure of FE-MCFormer and FE-MCFormer-s.}
\label{tabletrans}
\resizebox{\linewidth}{!}{%
\begin{tblr}{
  row{1} = {c},
  cell{1}{1} = {c=4}{},
  cell{1}{5} = {c=4}{},
  cell{5}{1} = {r=2}{},
  cell{5}{4} = {r=2}{},
  cell{5}{5} = {r=2}{},
  cell{5}{8} = {r=2}{},
  cell{7}{1} = {r=2}{},
  cell{7}{4} = {r=2}{},
  cell{7}{5} = {r=2}{},
  cell{7}{8} = {r=2}{},
  cell{9}{1} = {r=2}{},
  cell{9}{4} = {r=2}{},
  cell{9}{5} = {r=2}{},
  cell{9}{8} = {r=2}{},
  cell{11}{1} = {r=2}{},
  cell{11}{5} = {r=2}{},
  cell{13}{1} = {r=2}{},
  cell{13}{4} = {r=2}{},
  cell{13}{5} = {r=2}{},
  cell{13}{8} = {r=2}{},
  cell{15}{1} = {r=2}{},
  cell{15}{4} = {r=2}{},
  cell{15}{5} = {r=2}{},
  cell{15}{8} = {r=2}{},
  cell{17}{1} = {r=2}{},
  cell{17}{4} = {r=2}{},
  cell{17}{5} = {r=2}{},
  cell{17}{8} = {r=2}{},
  cell{19}{1} = {r=2}{},
  cell{19}{4} = {r=2}{},
  cell{19}{5} = {r=2}{},
  cell{19}{8} = {r=2}{},
  cell{21}{1} = {r=3}{},
  cell{21}{5} = {r=3}{},
  hline{1-5,7,9,11,13,15,17,19,21,24} = {-}{},
  hline{6,8,10,14,16,18,20} = {2-3,6-7}{},
  hline{12} = {2-4,6-7}{},
  hline{22-23} = {2-4,6-8}{},
}
FE-MCFormer        &                &                                         &                & FE-MCFormer-s       &                &                                         &                \\
Module             & Network Layer  & Parameter                               & Output shape   & Module              & Network Layer  & Parameter                               & Output shape   \\
                   & Input sequence &                                         & $1\times2048$  &                     & Input sequence &                                         & $1\times2048$  \\
Embedding          & FALL          & {kernel = 63,\\stride = 1}              & $32\times2048$ & Embedding           & FALL          & {kernel = 63,\\stride = 1}              & $16\times2048$ \\
Distilling          & Conv layer     & {kernel = 64,\\stride = 2}              & $32\times496$  & Distilling           & Conv layer     & {kernel = 64,\\stride = 2}              & $16\times496$  \\
                   & Pooling        & {kernel = 3,\\stride = 2}               &                &                     & Pooling        & {kernel = 3,\\stride = 2}               &                \\
{MSTFF\\block \#1} & $2\times$MSCAL & {B1 size = 5, B2 size = 3,\\stride = 1} & $32\times496$~ & {MSTFF\\block \#1 } & $1\times$MSCAL & {B1 size = 5, B2 size = 3,\\stride = 1} & $16\times496 $ \\
                   & $2\times$TFFN  & $\gamma = 0.1$                          &                &                     & $1\times$TFFN  & $\gamma = 0.1$                          &                \\
Distilling         & Conv layer     & {kernel = 5,\\stride = 1}               & $64\times247$  & Distilling          & Conv layer     & {kernel = 5,\\stride = 1}               & $32\times247$  \\
                   & Pooling        & {kernel = 3,\\stride = 2}               &                &                     & Pooling        & {kernel = 3,\\stride = 2}               &                \\
{MSTFF\\block \#2} & $2\times$MSCAL & {B1 size = 5, B2 size = 3,\\stride = 1} & $64\times247$  & {MSTFF\\block \#2 } & $1\times$MSCAL & {B1 size = 5, B2 size = 3,\\stride = 1} & $32\times247$  \\
                   & $2\times$TFFN  & $\gamma = 0.1$                          & $64\times247$  &                     & $1\times$TFFN  & $\gamma = 0.1$                          & $32\times247$  \\
Distilling         & Conv layer     & {kernel = 5,\\stride = 1}               & $128\times123$ & Distilling          & Conv layer     & {kernel = 5,\\stride = 1}               & $64\times123$  \\
                   & Pooling        & {kernel size = 3,\\stride = 2}          &                &                     & Pooling        & {kernel size = 3,\\stride = 2}          &                \\
{MSTFF\\block \#3} & $2\times$MSCAL & {B1 size = 5, B2 size = 3,\\stride = 1} & $128\times123$ & {MSTFF\\block \#3}  & $1\times$MSCAL & {B1 size = 5, B2 size = 3,\\stride = 1} & $64\times123$  \\
                   & $2\times$TFFN  & $\gamma = 0.1$                          &                &                     & $1\times$TFFN  & $\gamma = 0.1$                          &                \\
Distilling         & Conv layer     & {kernel = 5,\\stride=1}                 & $256\times61$  & Distilling          & Conv layer     & {kernel = 5,\\stride=1}                 & $128\times61$  \\
                   & Pooling        & {kernel = 3,\\stride = 2}               &                &                     & Pooling        & {kernel = 3,\\stride = 2}               &                \\
{MSTFF\\block \#4} & $2\times$MSCAL & {B1 size = 5, B2 size = 3,\\stride = 1} & $256\times61$  & {MSTFF\\block \#4}  & $1\times$MSCAL & {B1 size = 5, B2 size = 3,\\stride = 1} & $128\times61$  \\
                   & $2\times$TFFN  & $\gamma = 0.1$                          &                &                     & $1\times$TFFN  & $\gamma = 0.1$                          &                \\
Classifier         & Flatten layer  & /                                       & 15616          & Classifier          & Flatten layer  & /                                       & 7808           \\
                   & Linear layer   & /                                       & 256            &                     & Linear layer   & /                                       & 256            \\
                   & Linear layer   & /                                       & $C$            &                     & Linear layer   & /                                       & $C$            
\end{tblr}
}
\end{table*}

\subsection{Experimental settings}

The code of FE-MCFormer is implemented using PyTorch framework on a computer with an Intel(R) Xeon(R) Gold 6226R CPU and a RTX 4090 GPU. For training, FE-MCFormer uses cross-entropy as the loss function and Adam as the optimizer. The learning rate and batch size are set to 0.001 and 64, respectively.

The following max-min normalization operation is applied to mitigate the effects of various magnitudes of the measured values on the modeling accuracy:

\begin{flalign}
\tilde{x} = \frac{x-x_{min}}{x_{max}-x_{min}},
\label{eq-norm}
\end{flalign}

\noindent where \(\tilde{x}\) denotes the normalized value of $x$; \(x_{max}\) and \(x_{min}\) are the maximum and minimum values, respectively. Accuracy is used to evaluate fault diagnosis performance and is defined as:

\begin{flalign}
Acc = \frac{CP}{CP+IP} \times 100\%,
\end{flalign}

\noindent where \(CP\) indicates the correctly classified samples, and \(IP\) refers to incorrectly classified samples. To evaluate the performance of FE-MCFormer under noise, Gaussian white noise is added to the raw mechanical signals, resulting in signals with different SNRs, defined as follows:

\begin{flalign}
 SNR = 10\lg_{}{} \left ( \frac{P_s}{P_n}  \right ),  
\end{flalign}

\begin{flalign}
P_s = \frac{1}{N}\sum_{n=1}^{N}|v(n)|^2,
\end{flalign}

\noindent where \(P_s\) and \(P_n\) denote the power of the raw signal and the added noise, and \(v(n)\) represents the raw signal. In this experiment, the SNR is set in the range from \(-10\) dB to \(-2\) dB. Each group of experiments was repeated five times, with 200 epochs assigned to each experiment. Table \ref{tabletrans} shows the network structure of FE-MCFormer-s and FE-MCFormer, where FE-MCFormer-s is the lightweight version of the proposed method with fewer parameters and FE-MCFormer is the standard version with standard parameters. In the \(l\)-th MSTFF block, \(B1\) and \(B2\) represent convolution kernels of size \(1\times5\) and \(1\times3\) in Eq. (8), respectively; \(\gamma=0.1\) denotes the scaling factor \(\gamma\) in Eq. (5) and Eq. (14) is set to 0.1; \(C\) represents the number of classes in each dataset.

\subsection{Case 1: Rolling Bearing Dataset}
The bearing dataset \cite{lessmeier2016condition} from Paderborn University (PU) is employed to validate the FE-MCFormer under strong noise conditions.

\subsubsection{Dataset description}

\begin{table}
\centering
\caption{Description of PU dataset (Case 1)}
\label{table1-pudata}
\resizebox{\linewidth}{!}{%
\begin{tblr}{
  cells = {c,t},
  hlines,
  vlines,
}
Category & Bearing & Damage     & Level & Method               & Label    \\
0        & K001    & Normal     & /     & /                    & $H_1$    \\
1        & K002    & Normal     & /     & /                    & $H_2$    \\
2        & K003    & Normal     & /     & /                    & $H_3$    \\
3        & K004    & Normal     & /     & /                    & $H_4$    \\
4        & K005    & Normal     & /     & /                    & $H_5$    \\
5        & K006    & Normal     & /     & /                    & $H_6$    \\
6        & KA01    & Outer ring & 1     & EDM damage~ ~               & $H_7$    \\
7        & KA03    & Outer ring & 2     & electric engraver~ ~ & $H_8$    \\
8        & KA05    & Outer ring & 1     & EDM damage~ ~ & $H_9$    \\
9        & KA06    & Outer ring & 2     & electric engraver~ ~ & $H_{10}$ \\
10       & KA07    & Outer ring & 1     & drilling~ ~          & $H_{11}$ \\
11       & KA08    & Outer ring & 2     & drilling~ ~          & $H_{12}$ \\
12       & KA09    & Outer ring & 2     & drilling~ ~          & $H_{13}$ \\
13       & KI01    & Inner ring & 1     & EDM damage                  & $H_{14}$ \\
14       & KI03    & Inner ring & 1     & electric engraver~ ~ & $H_{15}$ \\
15       & KI05    & Inner ring & 1     & electric engraver~ ~ & $H_{16}$ \\
16       & KI07    & Inner ring & 2     & electric engraver~ ~ & $H_{17}$ \\
17       & KI08    & Inner ring & 2     & electric engraver~ ~ & $H_{18}$ 
\end{tblr}
}
\end{table}

\begin{figure}
    \centering
    \includegraphics[width=1\linewidth]{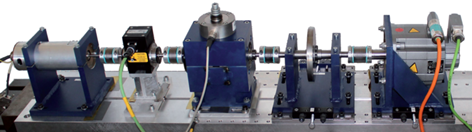}
    \caption{Rolling bearing test rig. (Case 1)}
    \label{fig.2}
\end{figure}

\begin{table*}
\centering
\caption{Test results of various data-driven methods on PU dataset in Case 1 (\%)}
\label{table2}
\resizebox{\linewidth}{!}{%
\begin{tblr}{
  row{2} = {c},
  cell{1}{1} = {r=2}{c},
  cell{1}{2} = {c},
  cell{1}{3} = {c},
  cell{1}{4} = {c},
  cell{1}{5} = {c},
  cell{1}{6} = {c},
  cell{1}{7} = {r=2}{},
  cell{1}{8} = {r=2}{},
  cell{4}{7} = {font=\bfseries},
  cell{4}{8} = {font=\bfseries},
  cell{10}{1} = {font=\bfseries},
  cell{10}{2} = {font=\bfseries},
  cell{10}{3} = {font=\bfseries},
  cell{10}{4} = {font=\bfseries},
  cell{10}{5} = {font=\bfseries},
  cell{10}{6} = {font=\bfseries},
  cell{11}{1} = {font=\bfseries},
  cell{11}{2} = {font=\bfseries},
  cell{11}{3} = {font=\bfseries},
  cell{11}{4} = {font=\bfseries},
  cell{11}{5} = {font=\bfseries},
  cell{11}{6} = {font=\bfseries},
  hline{1,3,12} = {-}{},
  hline{2} = {2-6}{},
}
Methods             & SNR=-10        & SNR=-8         & SNR=-6         & SNR=-4         & SNR=-2         & Params(M) & FLOPs(M)  \\
                    & Acc            & Acc            & Acc            & Acc            & Acc            &           &           \\
MSCNN-LSTM          & 50.17$\pm$2.17 & 61.90$\pm$5.66 & 75.39$\pm$5.09 & 85.65$\pm$1.11 & 91.13$\pm$1.20 & 0.09      & 57.37     \\
WDCNN               & 52.81$\pm$1.53 & 65.23$\pm$0.72 & 75.70$\pm$1.51 & 84.23$\pm$1.12 & 90.01$\pm$0.67 & 0.04      & 0.65      \\
ResNet50            & 58.85$\pm$0.47 & 72.90$\pm$0.62 & 83.86$\pm$0.54 & 90.57$\pm$0.47 & 96.49$\pm$0.34 & 15.99     & 1610.34   \\
DenseNet             & 60.81$\pm$0.33 & 74.11$\pm$0.72 & 84.15$\pm$0.39 & 90.31$\pm$0.92 & 94.15$\pm$0.59 & 3.80~     & 526.17    \\
Vanilla Transformer & 43.43$\pm$2.16 & 58.20$\pm$0.43 & 72.29$\pm$0.25 & 83.72$\pm$0.21 & 92.43$\pm$0.58 & 3.19      & 266.95    \\
Li-convformer       & 58.98$\pm$0.34 & 75.34$\pm$0.68 & 86.12$\pm$0.32 & 93.65$\pm$0.33 & 97.29$\pm$0.38 & 0.32~     & 28.83~ ~ \\
MCSwin-T            & 55.49$\pm$0.46 & 70.16$\pm$0.10 & 82.34$\pm$0.26 & 89.74$\pm$0.19 & 96.14$\pm$0.25 & 1.94      & 453.66    \\
FE-MCFormer-s       & 70.44$\pm$1.26 & 81.70$\pm$1.28 & 91.90$\pm$0.64 & 96.16$\pm$0.44 & 98.42$\pm$0.15 & 2.43~     & 54.60~ ~  \\
FE-MCFormer         & 72.99$\pm$0.54 & 86.31$\pm$0.51 & 94.74$\pm$0.17 & 98.01$\pm$0.14 & 99.11$\pm$0.18 & 7.24      & 338.95    
\end{tblr}
}
\end{table*}

\begin{figure}
    \centering
    \includegraphics[width=1\linewidth]{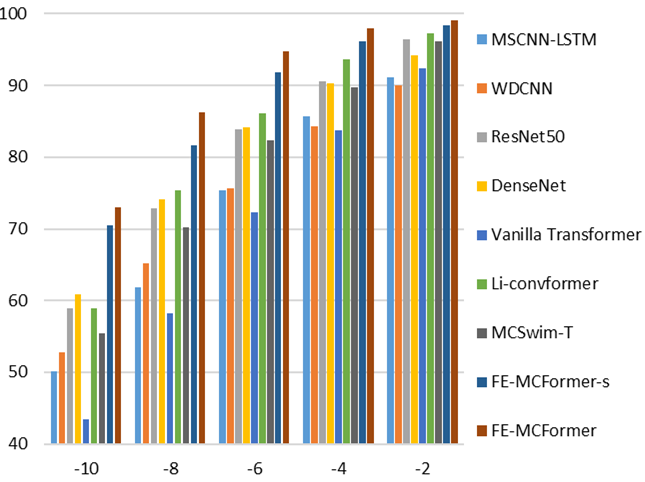}
    \caption{Test results of rolling bearing dataset in Case 1.}
    \label{fig.pu_acc}
\end{figure}

As shown in Fig. \ref{fig.2}, the rolling bearing test bench comprises an induction motor, a load motor, a measurement shaft, and a control panel. A deep groove ball bearing is selected as the test object. The experiment is conducted with the induction motor operating at 1500 rpm under a torque load of 0.7 Nm, with a sampling frequency of 64 kHz. A total of 18 bearing states are employed, including 6 normal states, 7 outer-ring faults, and 5 inner-ring faults, as detailed in Table \ref{table1-pudata}. For each health state, the raw vibration signal was segmented sequentially into 1000 consecutive samples, resulting in a total of 18 × 1000 = 18,000 samples, with each sample containing 2048 vibration points. To avoid data leakage, the first 80\% of the sequentially segmented samples were used for model training, while the remaining 20\% were reserved for testing.

\subsubsection{Overall Diagnostic Performance}\label{pu result}

To assess the effectiveness of FE-MCFormer, extensive experiments are conducted under varying noise conditions, with SNR ranging from \(- 10\) dB to \(- 2\) dB. Seven approaches, namely MSCNN-LSTM \cite{chen2021bearing}, Vanilla Transformer\cite{li_variational_2024}, WDCNN, ResNet50 \cite{zhao2020deep}, DenseNet, MCSwin-T \cite{chen2022multi} and Li-convformer \cite{yan_liconvformer_2024}, are employed for comparison with FE-MCFormer. From Table \ref{table2} and Fig. \ref{fig.pu_acc}, it can be observed that the proposed method consistently outperforms existing data-driven baselines across all noise levels in terms of diagnostic performance.

\begin{figure*}[!t]
    \centering
    \includegraphics[
        width=0.95\textwidth,
        height=0.85\textheight,
        keepaspectratio
    ]{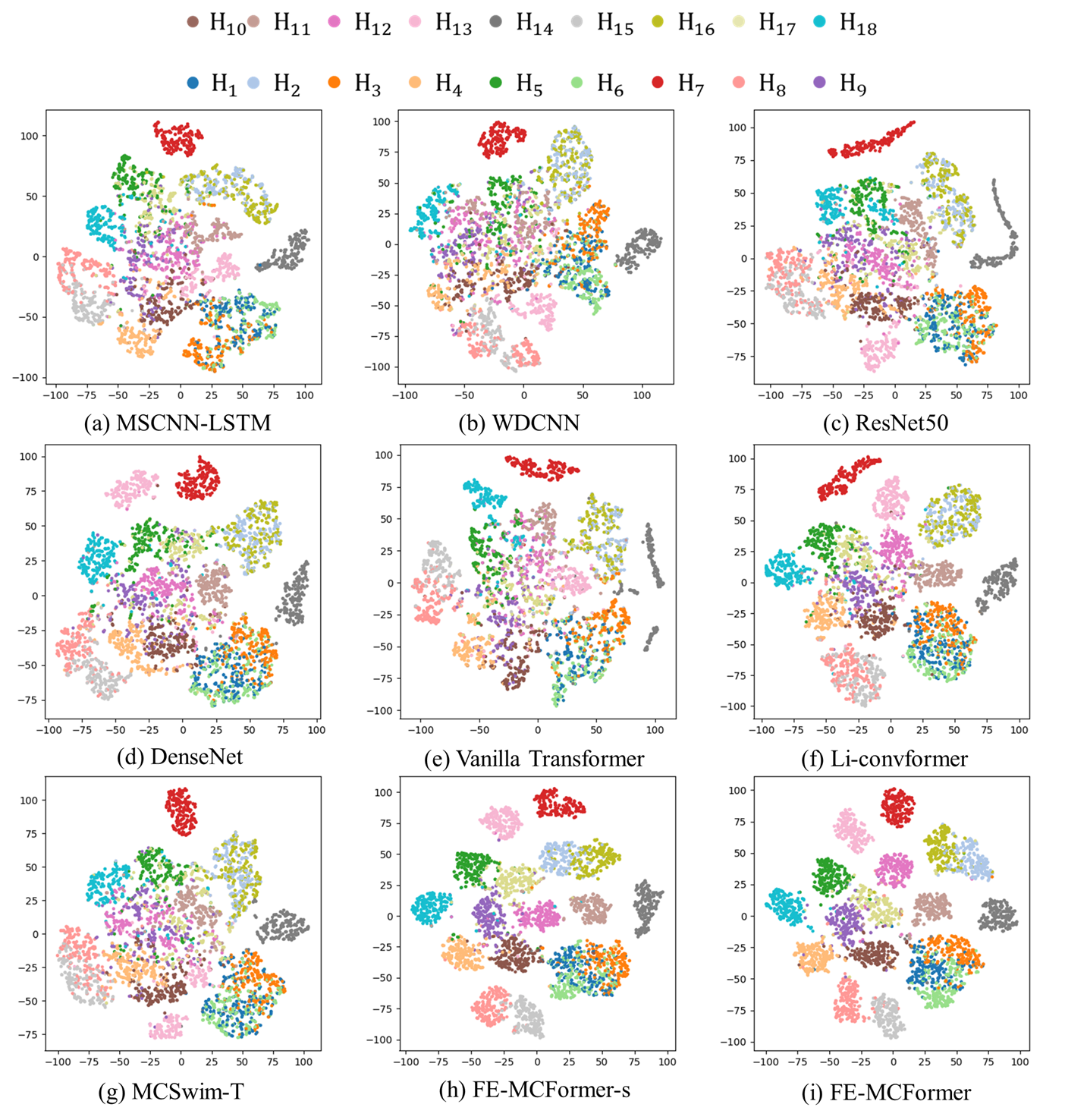}
    \caption{Results of 2D t-SNE visualization on the rolling bearing dataset under SNR = $-6$ dB in Case 1.}
    \label{fig.pu_tsne}
\end{figure*}

Specifically, FE-MCFormer and FE-MCFormer-s achieve superior performance across all noise scenarios, significantly outperforming both CNN-based and transformer-based methods. The proposed network architecture maintains high accuracy even under extremely noisy environments, further illustrating its strong robustness. Fig. \ref{fig.pu_tsne} shows the 2D T-SNE visualization results of various methods under SNR = \(-6\) dB, the results show dense clustering of different fault types with high classification accuracy, which further illustrates the feature discrimination capability of FE-MCFormer in complex industrial environments. These results indicate that FE-MCFormer architecture maintains a robust feature extraction capability even under severe noise interference, which is critical for industrial condition monitoring scenarios.

\subsubsection{Comparative Analysis}

MSCNN-LSTM, as a state-of-the-art multiscale convolutional network, exhibits outstanding performance under weak noise conditions, reaching 91.13\% mean accuracy at the SNR of \(-2\) dB. However, as the noise level increases from \(-2\) dB to \(-6\) dB, its accuracy drops significantly. In contrast, standard FE-MCFormer achieves diagnostic accuracy of 99.11\%, 98.01\%, and 94.74\% under the SNR from \(-2\) dB to \(-6\) dB. Numerical results indicate that, in noisy environments, the multiscale kernels in the multiscale convolutional neural networks (MSCNN) tend to amplify both high-frequency noise patterns and key local features, resulting in redundant and conflicting weight distributions. This makes it difficult for the network to focus on critical fault features, resulting in degraded performance in the FD of rotating machines. In contrast, the proposed FE-MCFormer leverages FALL and TFFN to suppress irrelevant frequency components and employs MSCAL to strengthen local multiscale feature representations, thereby achieving superior performance over MSCNNs.

The WDCNN uses large kernels to capture global dependencies, thus reducing noise interference. However, due to the limitation of the convolution kernel, the model struggles to fully capture long-range dependencies over the entire signal sequence. More specifically, as illustrated in Table \ref{table2}, the mean accuracies of WDCNN under SNR \(=\) \(-2\), \(-4\), \(-6\), \(-8\), \(-10\) are 90.01\%, 84.23\%, 75.70\%, 65.23\% and 52.81\%. Compared with standard FE-MCFormer and FE-MCFormer-s, FE-MCFormer consistently achieves higher accuracy under all noise conditions, indicating the effectiveness of FE-MCFormer.  

Compared with ResNet50 and DenseNet, FE-MCFormer achieves higher diagnostic accuracy under all noise conditions. This improvement can be attributed to its time–frequency modeling strategy. In conventional CNN architectures such as ResNet50 and DenseNet, small convolutional kernels mainly capture local patterns within a limited receptive field, and their responses can be easily affected by noise-induced fluctuations under low-SNR conditions. In contrast, FE-MCFormer combines frequency adaptive learning with multiscale temporal feature extraction, enabling the model to suppress noise-dominated components while retaining fault-sensitive structures. As reported in Table \ref{table2}, the mean accuracies of the standard FE-MCFormer under SNR levels of \(-2\), \(-4\), \(-6\), \(-8\), and \(-10\) dB are 99.11\%, 98.01\%, 94.74\%, 86.31\%, and 72.99\%, respectively. These results are consistently higher than those of ResNet50, which obtains 96.49\%, 90.57\%, 83.86\%, 72.90\%, and 58.85\%, respectively. Compared with DenseNet, FE-MCFormer also achieves accuracy improvements of 4.96\%, 7.70\%, 10.59\%, 12.20\%, and 12.18\% across the same noise levels. It is also worth noting that FE-MCFormer maintains lower computational cost than these CNN baselines. Specifically, the computational costs of FE-MCFormer and FE-MCFormer-s are 338.95M and 54.60M FLOPs, respectively, which are lower than those of ResNet50 (1610.34M FLOPs) and DenseNet (526.17M FLOPs). These results indicate that the proposed framework achieves a favorable balance between noise-robust diagnostic performance and computational efficiency compared with representative CNN-based models.

Vanilla Transformer shows a dramatic decrease in diagnostic performance under strong noise. This is mainly attributed to the fact that the severe noise contaminates the entire signal sequence, which prevents the MHSA from effectively modeling global features. The proposed FE-MCFormer utilizes FALL and TFFN to suppress irrelevant information and capture global dependencies. The MSCAL and distillation mechanism are further introduced to precisely extract the key local features, thus achieving higher performance.

\begin{figure*}
    \centering
    \includegraphics[width=0.95\textwidth,
        height=0.85\textheight,
        keepaspectratio]{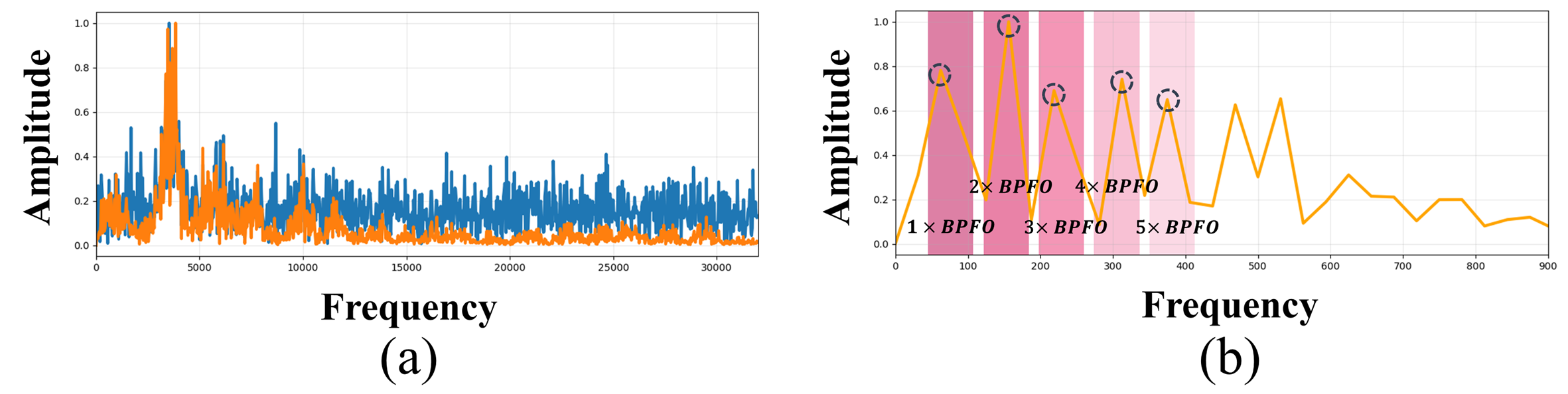}
    \caption{Reconstruction results of the bearing dataset (Case 1) under SNR = \(-4\) dB, KA01, outer ring fault. (a) Frequency spectrum analysis. (b) Envelope spectrum analysis.}
    \label{fig.pu_ka01}
\end{figure*}

\begin{figure*}
    \centering
    \includegraphics[width=0.95\textwidth,
        height=0.85\textheight,
        keepaspectratio]{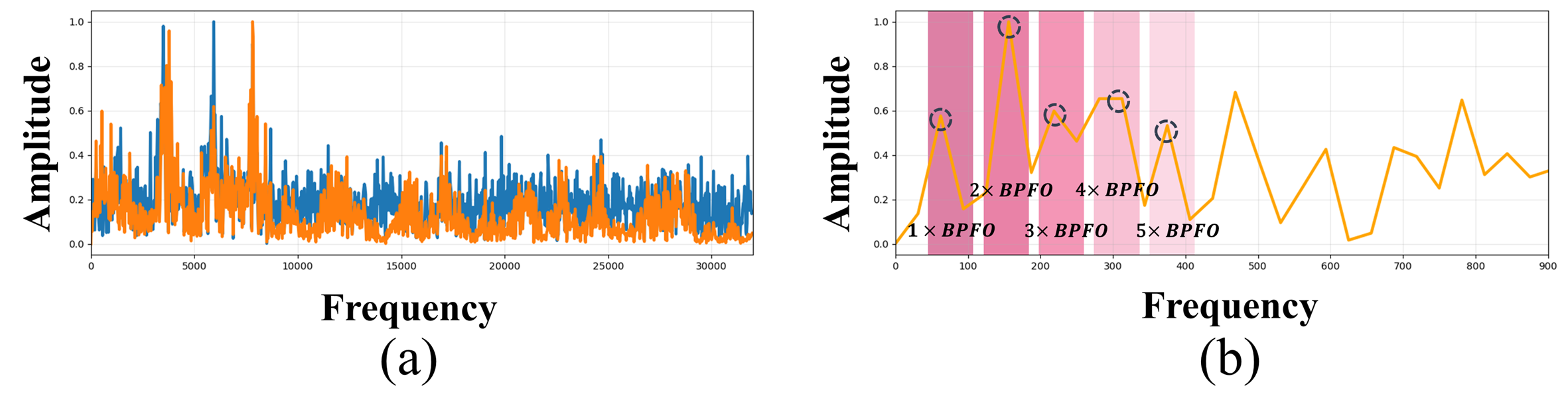}
    \caption{Reconstruction results of the bearing dataset (Case 1) under SNR = \(-4\) dB, KA03, outer ring fault. (a) Frequency spectrum analysis. (b) Envelope spectrum analysis.}
    \label{fig.pu_ka03}
\end{figure*}

\begin{figure*}
    \centering
    \includegraphics[width=0.95\textwidth,
        height=0.85\textheight,
        keepaspectratio]{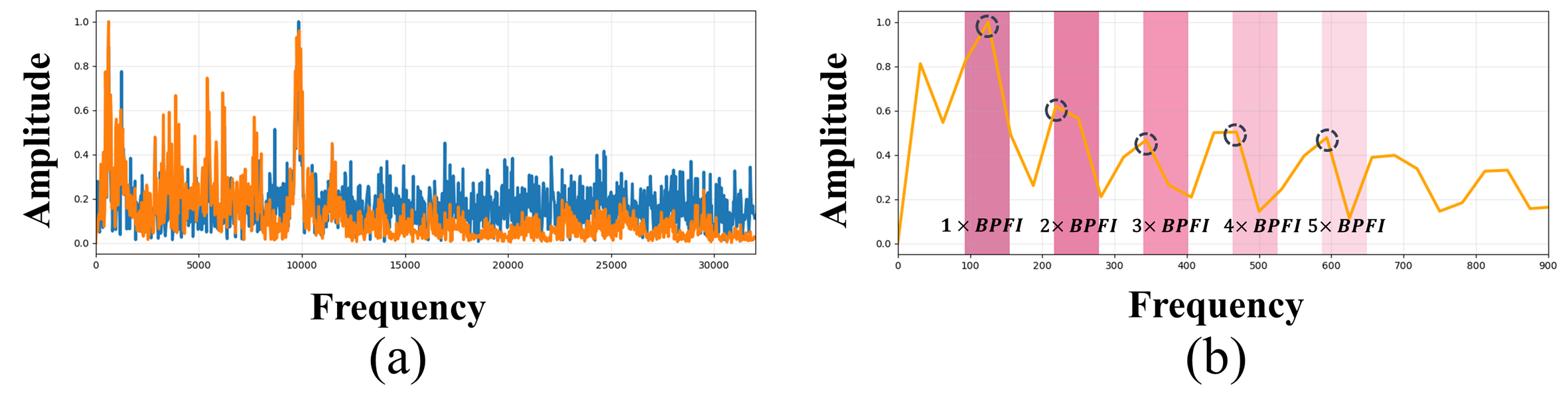}
    \caption{Reconstruction results of the bearing dataset (Case 1) under SNR = \(-4\) dB, KI01, inner ring fault. (a) Frequency spectrum analysis. (b) Envelope spectrum analysis.}
    \label{fig.pu_ki01}
\end{figure*}

\begin{figure*}
    \centering
    \includegraphics[width=0.95\textwidth,
        height=0.85\textheight,
        keepaspectratio]{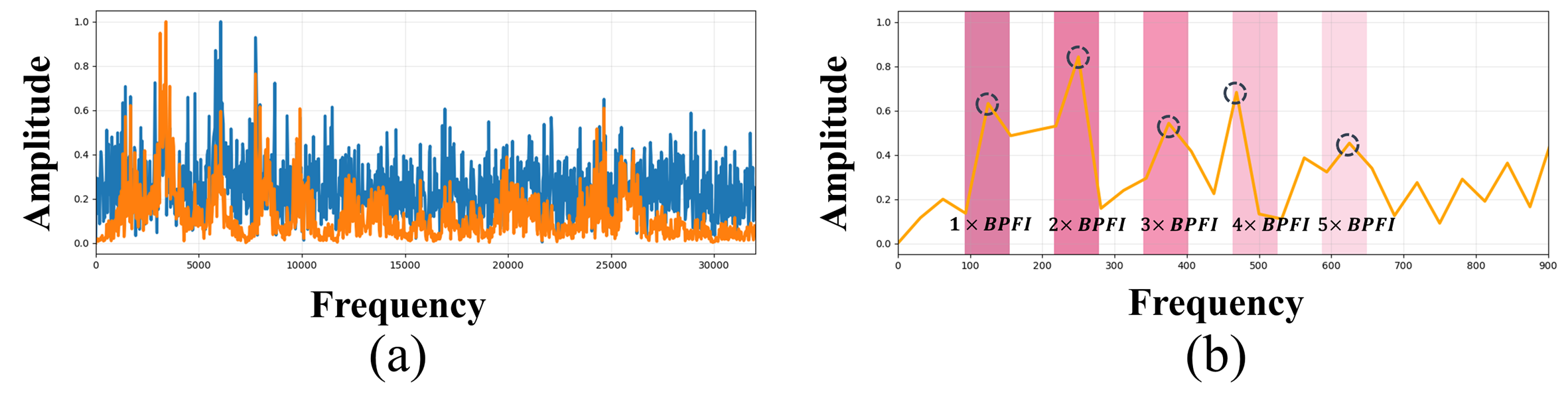}
    \caption{Reconstruction results of the bearing dataset (Case 1) under SNR = \(-4\) dB, KI07, inner ring fault. (a) Frequency spectrum analysis. (b) Envelope spectrum analysis.}
    \label{fig.pu_ki07}
\end{figure*}

\begin{table*}
\centering
\caption{Results of ablation study on PU dataset in Case 1}
\label{table6}
\resizebox{\linewidth}{!}{%
\begin{tblr}{
  row{2} = {c},
  cell{1}{1} = {r=2}{c},
  cell{1}{2} = {c},
  cell{1}{3} = {c},
  cell{1}{4} = {c},
  cell{1}{5} = {c},
  cell{1}{6} = {c},
  cell{1}{7} = {r=2}{},
  cell{1}{8} = {r=2}{},
  cell{3}{2} = {font=\bfseries},
  cell{3}{3} = {font=\bfseries},
  cell{3}{4} = {font=\bfseries},
  cell{3}{5} = {font=\bfseries},
  cell{3}{6} = {font=\bfseries},
  hline{1,3,7} = {-}{},
  hline{2} = {2-6}{},
}
Methods     & SNR=-10        & SNR=-8         & SNR=-6         & SNR=-4          & SNR=-2         & Params(M) & FLOPs(M) \\
            & Acc            & Acc            & Acc            & Acc             & Acc            &           &          \\
FE-MCFormer & 72.99$\pm$0.54 & 86.31$\pm$0.51 & 94.74$\pm$0.17 & 98.01$\pm$0.14  & 99.11$\pm$0.18 & 7.24      & 338.95   \\
Non-MSA     & 70.78$\pm$1.09 & 82.59$\pm$1.34 & 91.61$\pm$0.75 & 95.80$\pm$0.51 & 98.08$\pm$0.31 & 7.24      & 263.50   \\
Non-FAL-Head     & 69.77$\pm$0.81 & 81.28$\pm$0.33 & 90.99$\pm$0.74 & 95.95$\pm$0.58  & 98.22$\pm$0.40 & 6.98      & 338.07   \\
Non-FAL-Body   & 70.01$\pm$0.74 & 83.40$\pm$0.86 & 91.83$\pm$0.59 & 96.51$\pm$0.27  & 98.54$\pm$0.97 & 7.11      & 271.71   
\end{tblr}
}
\end{table*}

\begin{figure*}
    \centering
    \includegraphics[width=1\linewidth]{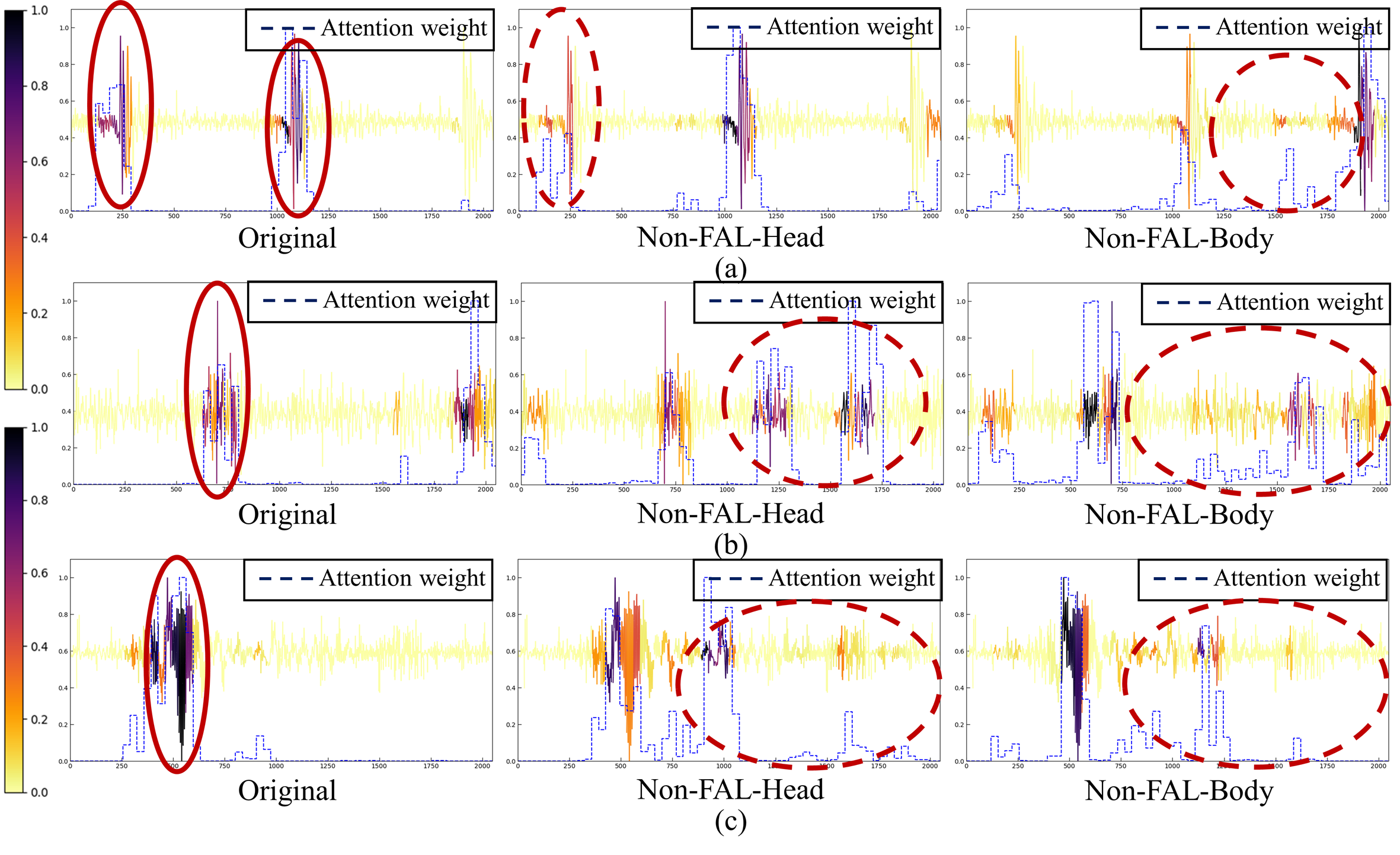}
    \caption{Attention heat map of FE-MCFormer under SNR = \(-8\) dB: (a) KA01, outer ring. (b) KA07, outer ring. (c) KI01, inner ring.}
    \label{fig.pu_abladation}
\end{figure*}

Li-convformer, as a state-of-the-art lightweight convformer-based fault diagnosis model, achieves competitive performance under relatively weak noise conditions, with mean accuracies of 97.29\%, 93.65\%, and 86.12\% at SNR levels of \(-2\), \(-4\), and \(-6\) dB, respectively. However, its diagnostic performance degrades noticeably as the noise intensity increases. In comparison, FE-MCFormer-s obtains mean accuracies of 98.42\%, 96.16\%, and 91.90\% under the same SNR levels, indicating improved robustness under noisy conditions. Moreover, FE-MCFormer-s maintains stable diagnostic performance even under severe noise interference, suggesting that the frequency adaptive learning and time–frequency modeling strategy can better preserve fault-sensitive features in low-SNR environments. From the computational perspective, FE-MCFormer-s requires 54.60M FLOPs, which remains comparable to Li-convformer with 28.83M FLOPs. Although FE-MCFormer-s introduces a moderate increase in computational cost, it provides higher diagnostic accuracy across the investigated noise levels. These results suggest that FE-MCFormer-s achieves a favorable trade-off between noise-robust diagnostic performance and computational efficiency compared with the lightweight convformer baseline.

Compared with MCSwin-T, which relies on a fixed-length sliding-window attention mechanism, the proposed method employs multiscale local attention to aggregate fault-related information across different temporal receptive fields. This design enables the model to capture localized impulsive patterns with varying durations, which are commonly observed in rotating machinery vibration signals. In addition, FALL and TFFN are introduced to further perform frequency-domain feature aggregation, allowing the model to suppress noise-dominated components while preserving fault-sensitive spectral structures. As a result, the proposed framework exhibits improved robustness under noisy operating conditions.

\subsubsection{Interpretability Analysis}

To validate the frequency-domain interpretability of the proposed FE-MCFormer under severe noise conditions, the reconstructed signals in test dataset produced by FALL are visualized in both the frequency domain and envelope spectrum. For illustration, Fig. \ref{fig.pu_ka01}, Fig. \ref{fig.pu_ka03}, Fig. \ref{fig.pu_ki01}, and Fig. \ref{fig.pu_ki07} present the reconstruction results under SNR = \(-4\) dB, where the blue curve denotes the noisy signal, the orange curve represents the FALL-reconstructed signal, and the pink shaded bands indicate the frequency-resolution regions around the ball pass frequency of outer race (BPFO) or the ball pass frequency of inner race (BPFI) and their harmonics.

As shown in Fig. \ref{fig.pu_ka01}, Fig. \ref{fig.pu_ka03}, Fig. \ref{fig.pu_ki01}, and Fig. \ref{fig.pu_ki07}, the reconstructed spectra suppress part of the broadband high-frequency noise components while retaining several fault-sensitive spectral responses. More importantly, the envelope spectra of the reconstructed signals exhibit pronounced responses within the frequency-resolution bands around the theoretical fault characteristic frequencies and their harmonics. Specifically, for the outer-race faults KA01 and KA03 shown in Fig. \ref{fig.pu_ka01} and Fig. \ref{fig.pu_ka03}, the dominant responses are distributed around the BPFO-related bands, including approximately 76 Hz, 152 Hz, 228 Hz, 305 Hz, and their neighboring harmonic regions. For the inner-race faults KI01 and KI07 shown in Fig. \ref{fig.pu_ki01} and Fig. \ref{fig.pu_ki07}, clear responses can be observed around the BPFI-related bands, including approximately 123 Hz, 247 Hz, 371 Hz, 494 Hz, and their neighboring harmonic regions. These results indicate that FALL can effectively suppress noise-dominated spectral components while preserving physically meaningful fault-related periodic modulation features, thereby providing frequency-domain interpretability for the proposed model.

To illustrate the interpretability of the proposed method in the time domain, we visualize the attention heatmaps of the first MSA in Eq. (\ref{eq-y3}) in the fourth MSTFF block. Fig. \ref{fig.pu_abladation} shows the attention heatmaps of three typical fault states in test dataset under SNR = \(-8\) dB, where darker colors represent regions that receive more attention from the model. The attention heatmaps in Fig. \ref{fig.pu_abladation} demonstrate that the proposed model focuses on physically meaningful fault impulse regions, as shown in the red solid circles in the figure. The above results indicate that the proposed FE-MCFormer remains interpretable even in a high-noise environment. More specifically, small-kernel convolutions (e.g., \(1\times 3\) and \(1\times 5\))  exhibit strong sensitivity to local signal components. This feature enables the small-kernel convolutions to effectively capture impulsive segments caused by rotating machinery faults. The proposed MSA module first employs multi-branch small-kernel convolutions to capture local fault characteristics at different temporal scales. Subsequently, sparse activation functions and channel-interaction convolutions are introduced to enhance cross-channel information exchange and nonlinear feature representation. Finally, an attention mechanism is adopted to adaptively emphasize fault-sensitive features while suppressing irrelevant responses, thereby improving the discriminative capability and interpretability of the model.

\subsubsection{Ablation Study and Component Effectiveness}

\begin{figure*}
    \centering
    \includegraphics[width=1\linewidth]{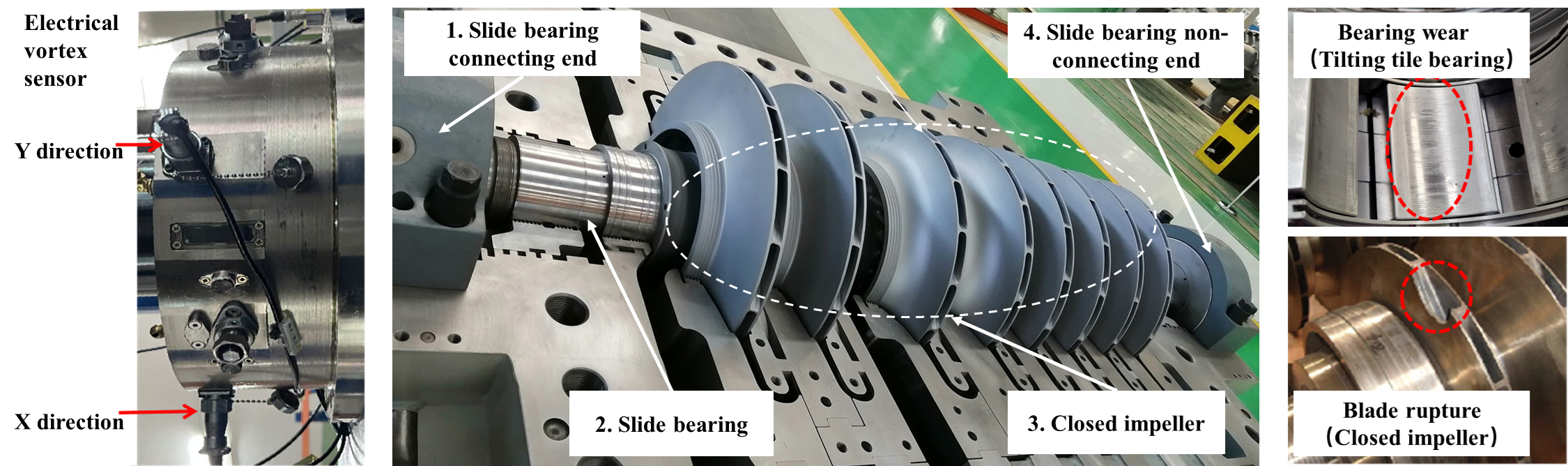}
    \caption{Illustration of the centrifugal compressor for methodology implementation (Case 2).}
    \label{fig.dlut_rig}
\end{figure*}

\begin{table}
\centering
\caption{Description of compressor dataset (Case 2)}
\label{table7}
\begin{tabular}{|c|c|c|c|} 
\hline
Category & Fault             & Label & Samples  \\ 
\hline
0        & Normal            & $C_1$ & 1000      \\ 
\hline
1        & Blade rupture     & $C_2$ & 1000      \\ 
\hline
2        & Block friction    & $C_3$ & 1000      \\ 
\hline
3        & Oil whirl & $C_4$ & 1000      \\ 
\hline
4        & Rotor imbalance  & $C_5$ & 1000      \\
\hline
\end{tabular}
\end{table}

Three variant models are constructed for validation on the PU dataset to assess the impact of different components in FE-MCFormer, namely Non-MSA, Non-FAL-Head and Non-FAL-Body. More specifically, Non-MSA indicates that the multiscale attention mechanism of Eq. (9) is not used; Non-FAL-Head denotes that the frequency adaptive learning (FAL) operations in FALL are not used; Non-FAL-Body indicates that the FAL in TFFN is not used. The ablation results shown in Table \ref{table6} confirm that each component contributes to the overall performance.

Table \ref{table6} demonstrates that the addition of MSA and FAL can effectively improve the robustness and accuracy of the model under strong noise conditions. More specifically, the introduction of MSA enables the model to extract key multiscale temporal features from the input representations, as illustrated in Table \ref{table6}. Compared with FE-MCFormer without the MSA mechanism, the FE-MCFormer achieves performance improvements of 2.21\%, 3.72\%, 3.13\%, 2.20\% and 1.03\% at the SNR levels of \(-10\), \(-8\), \(-6\), \(-4\) and \(-2\) dB, respectively. In the MSA, the \(1\times3\) and \(1\times5\) convolutional filters are introduced to extract multiscale features. The Hadamard product is used to activate the multiscale attention. Furthermore, the softmax function is applied to enhance the sparsity and interpretability of the attention maps.

FAL effectively suppresses noise-dominated spectra in the frequency domain. This mechanism not only improves the diagnostic accuracy of FE-MCFormer but also enhances the time-domain interpretability of the MSA module under severe noise conditions. The above conclusion can be further validated by visualizing the attention-weight distributions of MSA in different network variants.

As shown in Fig. \ref{fig.pu_abladation}, models without FAL exhibit significantly dispersed attention distributions under noisy conditions, particularly in the red dotted circle of Fig. \ref{fig.pu_abladation} (b) and (c). In fact, small-kernel convolutions (e.g., \(1\times3\) and \(1\times5\)) are highly effective in capturing local fault-related features under laboratory conditions with relatively low noise levels. However, as the noise intensity increases, fault-sensitive local information becomes progressively obscured, making it difficult for small-kernel convolutions to accurately identify discriminative patterns from complex signals.

Consequently, the diagnostic accuracy decreases significantly when FAL is removed, as shown in Table \ref{table6}. Meanwhile, the attention distributions learned by MSA become increasingly dispersed and susceptible to noise interference, resulting in reduced feature discrimination capability and poor interpretability. In contrast, the complete FE-MCFormer architecture exhibits highly concentrated attention around fault-related impulsive regions. This phenomenon suggests that FAL suppresses irrelevant spectral components and preserves fault-sensitive harmonic structures. As a result, MSA can focus more consistently on diagnostically informative temporal regions, thereby enhancing both the noise robustness and interpretability of FE-MCFormer under severe noise conditions.

\subsection{Case 2: Centrifugal Compressor Dataset}

\begin{table*}
\centering
\caption{Test results on the compressor dataset in Case 2 (\%)}
\label{dlut_acc}
\resizebox{\linewidth}{!}{%
\begin{tblr}{
  row{2} = {c},
  cell{1}{1} = {r=2}{c},
  cell{1}{2} = {c},
  cell{1}{3} = {c},
  cell{1}{4} = {c},
  cell{1}{5} = {c},
  cell{1}{6} = {c},
  cell{1}{7} = {r=2}{},
  cell{1}{8} = {r=2}{},
  cell{4}{7} = {font=\bfseries},
  cell{4}{8} = {font=\bfseries},
  cell{10}{1} = {font=\bfseries},
  cell{10}{2} = {font=\bfseries},
  cell{10}{3} = {font=\bfseries},
  cell{10}{4} = {font=\bfseries},
  cell{10}{5} = {font=\bfseries},
  cell{10}{6} = {font=\bfseries},
  cell{11}{1} = {font=\bfseries},
  cell{11}{2} = {font=\bfseries},
  cell{11}{3} = {font=\bfseries},
  cell{11}{4} = {font=\bfseries},
  cell{11}{5} = {font=\bfseries},
  cell{11}{6} = {font=\bfseries},
  hline{1,3,12} = {-}{},
  hline{2} = {2-6}{},
}
Methods             & SNR=-10        & SNR=-8         & SNR=-6         & SNR=-4         & SNR=-2         & Params(M) & FLOPs(M) \\
                    & Acc            & Acc            & Acc            & Acc            & Acc            &           &          \\
MSCNN-LSTM          & 63.69$\pm$2.04 & 72.05$\pm$2.29 & 74.81$\pm$1.75 & 88.70$\pm$0.63 & 90.54$\pm$2.17 & 0.092     & 57.37    \\
WDCNN               & 74.52$\pm$2.94 & 86.04$\pm$0.86 & 91.53$\pm$1.39 & 94.12$\pm$0.33 & 96.84$\pm$0.20 & 0.042     & 0.65     \\
ResNet50            & 75.90$\pm$0.87 & 84.01$\pm$0.49 & 89.90$\pm$0.37 & 94.05$\pm$0.22 & 94.99$\pm$0.50 & 15.99     & 1610.34  \\
DenseNet             & 81.56$\pm$1.22 & 88.39$\pm$0.41 & 93.00$\pm$0.49 & 95.32$\pm$0.44 & 96.66$\pm$0.61 & 3.80~     & 526.17   \\
Vanilla Transformer & 62.70$\pm$0.49 & 72.72$\pm$0.91 & 81.17$\pm$0.40 & 87.49$\pm$0.44 & 92.76$\pm$0.51 & 3.19      & 266.95   \\
Li-convformer       & 89.51$\pm$0.25 & 93.24$\pm$0.39 & 95.73$\pm$0.48 & 97.62$\pm$0.25 & 98.28$\pm$0.27 & 0.32~     & 28.83    \\
MCSwin-T             & 63.90$\pm$0.61 & 73.78$\pm$1.53 & 84.16$\pm$1.51 & 89.52$\pm$0.79 & 94.71$\pm$0.63 & 1.94      & 453.66   \\
FE-MCFormer-s       & 93.25$\pm$0.30 & 96.97$\pm$0.14 & 98.35$\pm$0.17 & 99.61$\pm$0.08 & 99.93$\pm$0.04 & 2.43~     & 54.60    \\
FE-MCFormer         & 93.48$\pm$1.18 & 96.63$\pm$0.15 & 98.46$\pm$0.18 & 99.71$\pm$0.07 & 99.98$\pm$0.04 & 7.24~     & 338.95   
\end{tblr}
}
\end{table*}

\subsubsection{Dataset description}

On the basis of fault diagnosis studies on small-scale rotating machinery, this work further extends the investigation to fault identification of large-scale turbomachinery. Large-scale compressors are critical turbomachines in the energy, petrochemical, metallurgical, and manufacturing industries, and their operating conditions are directly associated with production continuity, equipment safety, and energy utilization efficiency. Once faults occur, they may lead to increased vibration, degraded operating efficiency, and unexpected shutdowns, resulting in substantial economic losses and safety risks. Moreover, large-scale compressors usually operate in complex industrial environments, where the collected signals are easily affected by background noise, which poses a  challenge to the robustness of diagnostic models. To evaluate the practical applicability of FE-MCFormer, a real-world centrifugal compressor dataset is employed. The 7200 and 3300 eddy current probes are utilized to acquire the bearing displacement signals; the sampling frequency is set to 1024 Hz. In this study, the shaft displacement signals in the x-direction of the compressor bearing are collected under five operating conditions, including four fault states and one normal state, with 1000 samples for each condition, as summarized in Table \ref{table7}. Representative fault modes of the centrifugal compressor unit are illustrated in Fig. \ref{fig.dlut_rig}, such as block friction and blade rupture. Meanwhile, similar to the experiment in the PU dataset, the first 80\% of samples are used for model training, and the remaining 20\% are used for testing. Each sample contains 2048 points. All samples are normalized according to Eq. (\ref{eq-norm}).

\subsubsection{Comparative Analysis}

In this case, Gaussian white noise with SNR levels ranging from \(-10\) dB to \(-2\) dB is added to the raw signals to evaluate the robustness of the proposed method. Table \ref{dlut_acc} reports the diagnostic performance of various methods in this case. As shown in Table \ref{dlut_acc}, the proposed FE-MCFormer family achieves the best or highly competitive performance among the compared methods under all noisy conditions. More specifically, compared with MSCNN-LSTM and WDCNN, the proposed method provides clear accuracy improvements, particularly at low SNRs, indicating that frequency adaptive learning is beneficial for suppressing noise-dominated responses in compressor signals. Compared with ResNet50 and DenseNet, FE-MCFormer obtains higher accuracy with lower computational cost, while FE-MCFormer-s further reduces the FLOPs to 54.60M and still maintains competitive accuracy. In comparison with Transformer-based models such as the Vanilla Transformer and MCSwin-T, FE-MCFormer shows better robustness and avoids the high computational burden introduced by attention-based architectures. Moreover, compared with the strong lightweight convformer baseline Li-convformer, FE-MCFormer-s achieves higher accuracies across all SNR levels, with gains ranging from 1.65\% to 3.74\%. These results indicate that the proposed framework provides an effective trade-off between noise-robust fault diagnosis performance and computational efficiency for centrifugal compressor applications.

\subsubsection{Interpretability Analysis}

\begin{figure*}
\centering
\includegraphics[width=1\linewidth]{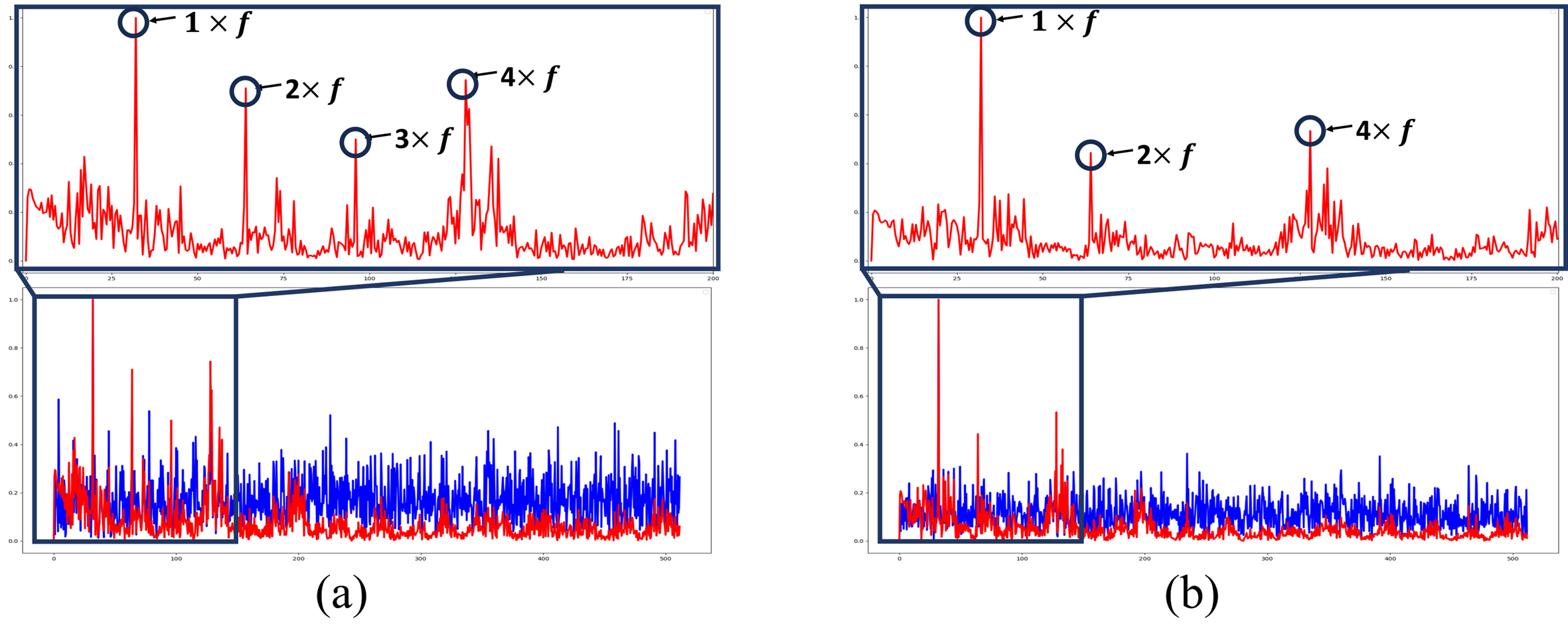}
\caption{Reconstruction results of the compressor dataset with four different types of faults described in Table \ref{table7} (Case 2) under SNR = \(-8\) dB, where the red curve denotes the reconstruction results, the blue curve denotes the original noisy signal. (a) Blade rupture. (b) Block friction.}
\label{fig.dlut_visualize_1}
\end{figure*}

\begin{figure*}
\centering
\includegraphics[width=1\linewidth]{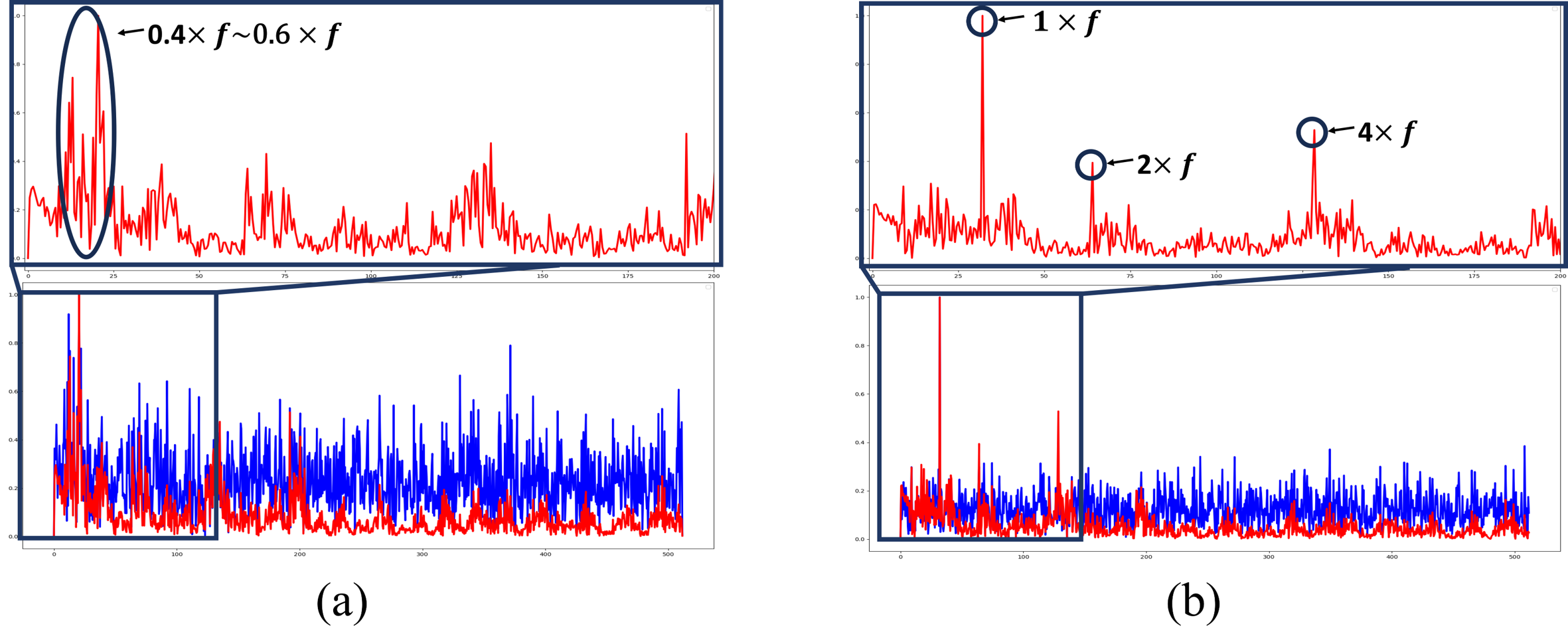}
\caption{Reconstruction results of the compressor dataset with four different types of faults described in Table \ref{table7} (Case 2) under SNR = \(-8\) dB, where the red curve denotes the reconstruction results, the blue curve denotes the original noisy signal. (a) Oil whirl. (b) Rotor imbalance.}
\label{fig.dlut_visualize_2}
\end{figure*}

It is worth noting that, unlike rolling-bearing faults, the abnormal trajectory of the shaft centerline caused by compressor failure does not result in obvious temporal impulsive patterns in the time domain. Therefore, this section mainly focuses on analyzing the interpretability of the model in the frequency domain. To demonstrate the practical interpretability and effectiveness of FE-MCFormer, we compare the normalized frequency spectra of the original and reconstructed signals of FALL in test dataset under SNR = \(-8\) dB, as illustrated in Fig. \ref{fig.dlut_visualize_1} and Fig. \ref{fig.dlut_visualize_2}, where the blue curves denote the original spectra, and the red ones correspond to the reconstructed signals; The horizontal axis represents frequency, and the vertical axis represents amplitude. Here, \(i\times f\) denotes the fault rotational frequency and its harmonics. 

It can be observed from Fig. \ref{fig.dlut_visualize_1} and Fig. \ref{fig.dlut_visualize_2} that FALL attenuates substantial high-frequency noise while preserving fault-sensitive spectral components and their harmonics. For instance, in the compressor blade rupture case shown in Fig. \ref{fig.dlut_visualize_1}(a), the signal reconstructed by FALL retains the dominant low-frequency components associated with the fault condition, including the rotational component at 32 Hz (\(1\times f\)) and its harmonics at 64 Hz (\(2\times f\)), 96 Hz (\(3\times f\)), and 128 Hz (\(4\times f\)).

For the block-friction case shown in Fig. \ref{fig.dlut_visualize_1}(b), pronounced spectral peaks can also be observed at the rotational frequency of 32 Hz (\(1\times f\)) and its harmonic components, such as 64 Hz (\(2\times f\)) and 128 Hz (\(4\times f\)). This is consistent with the nonlinear vibration characteristics commonly induced by rubbing-related faults. In contrast, for the oil-whirl case shown in Fig. \ref{fig.dlut_visualize_2}(a), particular attention should be paid to the preservation of sub-synchronous components (\(0.4\times f\sim0.6\times f\)), since oil-whirl phenomena are typically associated with frequency components below the rotational frequency. The retained rotational component may reflect the synchronous rotor response rather than the primary characteristic component of oil whirl.

Rotor imbalance is typically characterized by a dominant spectral component near the rotational frequency (\(1\times f\)), as illustrated in Fig. \ref{fig.dlut_visualize_2}(b). The preservation of these physically relevant and fault-sensitive spectral components across different fault conditions indicates that the proposed method can focus on meaningful frequency information, thereby improving the interpretability of the diagnostic process in addition to enhancing classification performance.

\section{Concluding Remarks} \label{Conclusion}
Interpretable fault diagnosis (FD) of rotating machinery becomes particularly challenging under strong noise conditions. Most deep learning (DL) methods fail to fully exploit the time–frequency features of input signals, resulting in poor performance under noisy conditions. To address this issue, this study proposes a frequency-enhanced multiscale transformer (FE-MCFormer) model for time–frequency interpretable FD of rotating machinery under severe noise conditions.

First, a frequency adaptive learning layer (FALL) is introduced to fully explore global frequency features while filtering out irrelevant information. Next, a multiscale time–frequency fusion (MSTFF) module is employed to extract multiscale time-domain information and global frequency-domain features. Additionally, a distillation layer is incorporated to expand the receptive field. Based on these modules, two novel deep learning architectures named FE-MCFormer and FE-MCFormer-s are developed based on the proposed improvements.

The effectiveness  and interpretability of FE-MCFormer are validated through case studies using two different datasets. Compared with seven DL methods, FE-MCFormer achieves improved diagnostic results and demonstrates strong noise robustness. For example, on the bearing dataset, FE-MCFormer achieves accuracies of 99.11\%, 98.01\%, 94.74\%, 86.31\% and 72.99\% at SNR levels of \(-2\), \(-4\), \(-6\), \(-8\) and \(-10\) dB, respectively. Meanwhile, on the compressor dataset, FE-MCFormer achieves accuracies of 99.98\%, 99.71\%, 98.46\%, 96.63\% and 93.48\% at SNR levels of \(-2\), \(-4\), \(-6\), \(-8\) and \(-10\) dB, respectively, higher than other DL methods, illustrating its strong potential for deployment in real-world industrial condition monitoring. Furthermore, comparative experiments are conducted to demonstrate the necessity of each component in FE-MCFormer. Interpretability analysis in both the time and frequency domains further reveals the internal mechanisms of the proposed FE-MCFormer. 

In future work, we will focus on expanding the application field and further enhancing the interpretability and noise robustness of both FALL and FE-MCFormer.

\section{Declarations}
\subsection{Competing Interests}
We would like to declare that the work described is original research that has not been published previously and is not under consideration for publication elsewhere, in whole or in part. We have no known competing financial interests or personal relationships that could have appeared to influence the work reported in this paper.

\subsection{Acknowledgment}
This work was supported in part by the National Natural Science Foundation in China under Grant No. U2541267, Liaoning Department of Science and Technology under Grant No. 2023JH1/10400084, and Dalian High-level Talent Innovation Program under Grant No. 2022RG10, which is highly appreciated.

\subsection{Author contribution}
Yuhan Yuan: Conceptualization, Data curation, Formal analysis, Investigation, Methodology, Software, Validation, Writing – original draft.

Xiaomo Jiang: Conceptualization, Funding acquisition, Project administration, Resources, Supervision, Writing – review \& editing.

Haibin Yang: Methodology, Validation.

Haixin Zhao: Validation, Supervision, Writing – review \& editing.

Shengbo Wang: Software, Validation.

Xueyu Cheng: Validation, Supervision, Writing – review \& editing.

Jigang Meng: Project administration, Resources.

\bibliographystyle{cas-model2-names}

\bibliography{cas-refs}

@article{zhang_semi-supervised_2023,
	title = {Semi-supervised fault diagnosis of gearbox based on feature pre-extraction mechanism and improved generative adversarial networks under limited labeled samples and noise environment},
	volume = {58},
	issn = {14740346},
	doi = {10.1016/j.aei.2023.102211},
	pages = {102211},
	journal = {Adv. Eng. Inform.},
	author = {Zhang, Lijie and Wang, Bin and Liang, Pengfei and Yuan, Xiaoming and Li, Na},
	urlyear = {2025-03-13},
	year = {2023},
	langid = {english},
}

@article{jiang_time-frequency_2023,
	title = {A time-frequency spectral amplitude modulation method and its applications in rolling bearing fault diagnosis},
	volume = {185},
	issn = {08883270},
	doi = {10.1016/j.ymssp.2022.109832},
	pages = {109832},
	journal = {Mech. Syst. Signal Process.},
	author = {Jiang, Zuhua and Zhang, Kun and Xiang, Ling and Yu, Gang and Xu, Yonggang},
	urlyear = {2025-03-13},
	year = {2023},
	langid = {english},
}

@article{chen_feature_2023,
	title = {Feature Extraction Based on Hierarchical Improved Envelope Spectrum Entropy for Rolling Bearing Fault Diagnosis},
	volume = {72},
	issn = {0018-9456, 1557-9662},
	doi = {10.1109/TIM.2023.3277938},
	pages = {1--12},
	journal = {{IEEE} Trans. Instrum. Meas.},
	author = {Chen, Zhixiang and Yang, Yang and He, Changbo and Liu, Yongbin and Liu, Xianzeng and Cao, Zheng},
	urlyear = {2025-03-13},
	year = {2023},
	langid = {english},
}

@article{kou_application_2020,
	title = {Application of {ICEEMDAN} Energy Entropy and {AFSA}-{SVM} for Fault Diagnosis of Hoist Sheave Bearing},
	volume = {22},
	issn = {1099-4300},
	doi = {10.3390/e22121347},
	pages = {1347},
	number = {12},
	journal = {Entropy},
	author = {Kou, Ziming and Yang, Fen and Wu, Juan and Li, Tengyu},
	urlyear = {2025-03-13},
	year = {2020},
	langid = {english},
}

@article{zhou_novel_2023,
	title = {A novel rolling bearing fault diagnosis method based on continuous hierarchical fractional range entropy},
	volume = {220},
	issn = {02632241},
	doi = {10.1016/j.measurement.2023.113395},
	pages = {113395},
	journal = {Measurement},
	author = {Zhou, Jie and Chen, Chuanhai and Guo, Jinyan and Wang, Liding and Liu, Zhifeng and Feng, Cong},
	urlyear = {2025-03-13},
	year = {2023},
	langid = {english},
}

@article{ren_meta-learning_2024,
	title = {Meta-Learning With Distributional Similarity Preference for Few-Shot Fault Diagnosis Under Varying Working Conditions},
	volume = {54},
	issn = {2168-2267, 2168-2275},
	doi = {10.1109/TCYB.2023.3338768},
	pages = {2746--2756},
	number = {5},
	journal = {{IEEE} Trans. Cybern.},
	author = {Ren, Chao and Jiang, Bin and Lu, Ningyun and Simani, Silvio and Gao, Furong},
	urlyear = {2025-03-13},
	year = {2024},
	langid = {english},
}

@article{jia_multiscale_2022,
	title = {Multiscale Residual Attention Convolutional Neural Network for Bearing Fault Diagnosis},
	volume = {71},
	issn = {0018-9456, 1557-9662},
	doi = {10.1109/TIM.2022.3196742},
	pages = {1--13},
	journal = {{IEEE} Trans. Instrum. Meas.},
	author = {Jia, Linshan and Chow, Tommy W. S. and Wang, Yu and Yuan, Yixuan},
	urlyear = {2025-03-15},
	year = {2022},
	langid = {english},

}

@inproceedings{lessmeier2016condition,
  title={Condition monitoring of bearing damage in electromechanical drive systems by using motor current signals of electric motors: A benchmark data set for data-driven classification},
  author={Lessmeier, Christian and Kimotho, James Kuria and Zimmer, Detmar and Sextro, Walter},
  booktitle={PHM society European conference},
  volume={3},
  year={2016}
}

@article{zhao2020deep,
title = {Deep learning algorithms for rotating machinery intelligent diagnosis: An open source benchmark study},
journal = {ISA Transactions},
volume = {107},
pages = {224-255},
year = {2020},
issn = {0019-0578},
doi = {https://doi.org/10.1016/j.isatra.2020.08.010},
url = {https://www.sciencedirect.com/science/article/pii/S0019057820303335},
author = {Zhibin Zhao and Tianfu Li and Jingyao Wu and Chuang Sun and Shibin Wang and Ruqiang Yan and Xuefeng Chen}
}

@article{chen2021bearing,
  title={Bearing fault diagnosis base on multi-scale CNN and LSTM model},
  author={Chen, Xiaohan and Zhang, Beike and Gao, Dong},
  journal={J. Intell. Manuf.},
  volume={32},
  number={4},
  pages={971--987},
  year={2021},
  publisher={Springer}
}

@article{li_variational_2024,
	title = {Variational {Attention}-{Based} {Interpretable} {Transformer} {Network} for {Rotary} {Machine} {Fault} {Diagnosis}},
	volume = {35},
	journal = {IEEE Trans. Neural Netw. Learning Syst.},
	author = {Li, Yasong and Zhou, Zheng and Sun, Chuang and Chen, Xuefeng and Yan, Ruqiang},
	month = may,
	year = {2024},
	publisher={IEEE},
	pages = {6180--6193},
}

@article{yan_liconvformer_2024,
	title = {{LiConvFormer}: A lightweight fault diagnosis framework using separable multiscale convolution and broadcast self-attention},
	volume = {237},
	doi = {10.1016/j.eswa.2023.121338},
	shorttitle = {{LiConvFormer}},
	pages = {121338},
	journal = {Expert Systems with Applications},
	author = {Yan, Shen and Shao, Haidong and Wang, Jie and Zheng, Xinyu and Liu, Bin},
	year = {2024},
}

@article{YUAN2025322,
title = {Remaining useful life prediction for the harmonic reducer of industrial robots via in-situ current signal and lightweight multiscale attention deep networks},
journal = {Journal of Manufacturing Systems},
volume = {83},
pages = {322-336},
year = {2025},
issn = {0278-6125},
doi = {https://doi.org/10.1016/j.jmsy.2025.09.008},
url = {https://www.sciencedirect.com/science/article/pii/S0278612525002353},
author = {Yuhan Yuan and Yanfeng Han and Ke Xiao and Zhongying Xu and Xiaomo Jiang}
}

@article{lei2020applications,
  title={Applications of machine learning to machine fault diagnosis: A review and roadmap},
  author={Lei, Yaguo and Yang, Bin and Jiang, Xinwei and Jia, Feng and Li, Naipeng and Nandi, Asoke K},
  journal={Mech. Syst. Signal Process.},
  volume={138},
  pages={106587},
  year={2020},
  publisher={Elsevier}
}

@article{ahmad2025consistency,
  title={Consistency-regularized-label-aware contrastive learning with uncertainty-aware periodic pseudo-labeling for machinery fault diagnosis under limited labeled data},
  author={Ahmad, Hassaan and Cheng, Wei and Wang, Wentao and Zhang, Shou and Nie, Zelin and Liu, Haoyu and Chen, Xuefeng},
  journal={Adv. Eng. Inform.},
  volume={68},
  pages={103656},
  year={2025},
  publisher={Elsevier}
}

@article{ZHAO2026104189,
title = {Physics modeling-driven interpretable data augmentation method for bearing fault diagnosis under imbalanced data},
journal = {Adv. Eng. Inform.},
volume = {70},
pages = {104189},
year = {2026},
issn = {1474-0346},
doi = {https://doi.org/10.1016/j.aei.2025.104189},
url = {https://www.sciencedirect.com/science/article/pii/S1474034625010821},
author = {Lijuan Zhao and Junyu Qi and Fei Wu and Yi Wang and Yi Qin},
}

@article{he2026hybrid,
  title={A hybrid cross-domain few-shot bearing fault diagnosis method combining multi-scale feature association and physical information},
  author={He, Changfu and Gai, Nanyan and Yan, Ke and Shao, Haidong},
  journal={Adv. Eng. Inform.},
  volume={69},
  pages={104077},
  year={2026},
  publisher={Elsevier}
}

@article{zhang2025wd,
  title={WD-KANTF: An interpretable intelligent fault diagnosis framework for rotating machinery under noise environments and small sample conditions},
  author={Zhang, Yazhou and Zhao, Xiaoqiang and Peng, Zhenrui and Xu, Rongrong and Chen, Peng},
  journal={Adv. Eng. Inform.},
  volume={66},
  pages={103452},
  year={2025},
  publisher={Elsevier}
}

@article{chen2022multi,
  title={Multi-channel calibrated transformer with shifted windows for few-shot fault diagnosis under sharp speed variation},
  author={Chen, Zhuohang and Chen, Jinglong and Liu, Shen and Feng, Yong and He, Shuilong and Xu, Enyong},
  journal={ISA transactions},
  volume={131},
  pages={501--515},
  year={2022},
  publisher={Elsevier}
}

@article{CHEN2022501,
title = {Multi-channel Calibrated Transformer with Shifted Windows for few-shot fault diagnosis under sharp speed variation},
journal = {ISA Transactions},
volume = {131},
pages = {501-515},
year = {2022},
issn = {0019-0578},
author = {Zhuohang Chen and Jinglong Chen and Shen Liu and Yong Feng and Shuilong He and Enyong Xu},
}

@article{LIAO2025111750,
title = {Classifier-guided neural blind deconvolution: A physics-informed denoising module for bearing fault diagnosis under noisy conditions},
journal = {Mechanical Systems and Signal Processing},
volume = {222},
pages = {111750},
year = {2025},
issn = {0888-3270},
doi = {https://doi.org/10.1016/j.ymssp.2024.111750},
url = {https://www.sciencedirect.com/science/article/pii/S0888327024006484},
author = {Jing-Xiao Liao and Chao He and Jipu Li and Jinwei Sun and Shiping Zhang and Xiaoge Zhang},
}

@article{CHEN2026130747,
title = {A multi-channel signal fault diagnosis method based on dynamic weighted data fusion and multi-scale feature enhancement},
journal = {Expert Systems with Applications},
volume = {306},
pages = {130747},
year = {2026},
issn = {0957-4174},
author = {Ke Chen and Feilong Zhou and Fangfang Zhang and Kunjie Yu and Duo Yang},
}

@article{WANG2025112889,
title = {An ensemble deep learning network based on 2D convolutional neural network and 1D LSTM with self-attention for bearing fault diagnosis},
journal = {Applied Soft Computing},
volume = {172},
pages = {112889},
year = {2025},
issn = {1568-4946},
author = {Liying Wang and Weiguo Zhao}
}

@article{LIU2025111271,
title = {Adaptive frequency attention-based interpretable Transformer network for few-shot fault diagnosis of rolling bearings},
journal = {Reliability Engineering \& System Safety},
volume = {263},
pages = {111271},
year = {2025},
issn = {0951-8320},
doi = {https://doi.org/10.1016/j.ress.2025.111271},
url = {https://www.sciencedirect.com/science/article/pii/S0951832025004727},
author = {Keying Liu and Yifan Li and Zhaoyang Cui and Guangdong Qi and Biao Wang}
}

@ARTICLE{10870139,
  author={Wang, Zhen and Han, Guangjie and Liu, Li and Wang, Feng and Zhu, Yuanyang},
  journal={IEEE Transactions on Instrumentation and Measurement}, 
  title={A Globally Interpretable Convolutional Neural Network Combining Bearing Semantics for Bearing Fault Diagnosis}, 
  year={2025},
  volume={74},
  number={},
  pages={1-13},
  doi={10.1109/TIM.2025.3538068}}

@article{CHEN2024102705,
title = {VKCNN: An interpretable variational kernel convolutional neural network for rolling bearing fault diagnosis},
journal = {Advanced Engineering Informatics},
volume = {62},
pages = {102705},
year = {2024},
issn = {1474-0346},
doi = {https://doi.org/10.1016/j.aei.2024.102705},
url = {https://www.sciencedirect.com/science/article/pii/S1474034624003537},
author = {Guangyi Chen and Gang Tang and Zhixiao Zhu},
}

@ARTICLE{10584499,
  author={Zhou, Qianyu and Tang, Jiong},
  journal={IEEE Internet of Things Journal}, 
  title={An Interpretable Parallel Spatial CNN-LSTM Architecture for Fault Diagnosis in Rotating Machinery}, 
  year={2024},
  volume={11},
  number={19},
  pages={31730-31744},
  doi={10.1109/JIOT.2024.3422969}}

\end{document}